\documentclass[longauth]{aa}

\usepackage{subfigure}
\usepackage{float}
\usepackage{txfonts}
\usepackage{graphicx}
\usepackage{natbib}
\usepackage{xspace}

\bibpunct{(}{)}{;}{a}{}{,}

\def\xirppi{\xi^s(r_p,\pi)}
\def\({\left(}
\def\){\right)}
\def\[{\left[}
\def\]{\right]}
\def\mhmpc{\,h^{-1}{\rm Mpc}}

\newcommand*\pdt{P_{\delta\theta}\(k\)}
\newcommand*\ptt{P_{\theta\theta}\(k\)}
\newcommand*\bb{b_{\rm{b}}\sigma_8}
\newcommand*\br{b_{\rm{r}}\sigma_8}

\newcommand{\dd}{\mathrm{d}}

\newcommand{\hmpc}{$\,h^{-1}$ Mpc\xspace}
\newcommand{\kms}{$\,{\rm km\, s^{-1}}$\xspace}

\def\bigstrutup{\vrule width0pt height0.3truecm depth0truecm}
\def\bigstrutdown{\vrule width0pt height0truecm depth0.16truecm}

\begin{document}

\title{The VIMOS Public Extragalactic Redshift Survey (VIPERS)\thanks{
    Based on observations collected at the European Southern
    Observatory, Cerro Paranal, Chile, using the Very Large Telescope
    under programs 182.A-0886 and partly 070.A-9007.  Also based on
    observations obtained with MegaPrime/MegaCam, a joint project of
    CFHT and CEA/DAPNIA, at the Canada-France-Hawaii Telescope (CFHT),
    which is operated by the National Research Council (NRC) of
    Canada, the Institut National des Sciences de l’Univers of the
    Centre National de la Recherche Scientifique (CNRS) of France, and
    the University of Hawaii. This work is based in part on data
    products produced at TERAPIX and the Canadian Astronomy Data
    Centre as part of the Canada-France-Hawaii Telescope Legacy
    Survey, a collaborative project of NRC and CNRS. The VIPERS web
    site is http://www.vipers.inaf.it/.}  }

\subtitle{
An unbiased estimate of the growth rate of structure at $\mathbf{\left<z\right>=0.85}$ using the clustering of luminous blue galaxies}

\titlerunning{Redshift-space distortions in VIPERS}

\author{
F.~G.~Mohammad\inst{\ref{brera},\ref{insubria},\ref{unimi}}
\and B.~R.~Granett\inst{\ref{brera},\ref{unimi}}                                                 
\and L.~Guzzo\inst{\ref{unimi},\ref{brera}} 
\and J.~Bel\inst{\ref{cpt},\ref{brera}}
\and E.~Branchini\inst{\ref{roma3},\ref{infn-roma3},\ref{oa-roma}}
\and S.~de la Torre\inst{\ref{lam}}
\and L.~Moscardini\inst{\ref{unibo},\ref{infn-bo},\ref{oabo}}
\and J.~A.~Peacock\inst{\ref{roe}}
%
%
\and M.~Bolzonella\inst{\ref{oabo}}      
\and B.~Garilli\inst{\ref{iasf-mi}}          
\and M.~Scodeggio\inst{\ref{iasf-mi}}       
%
%
\and U.~Abbas\inst{\ref{oa-to}}
\and C.~Adami\inst{\ref{lam}}
\and D.~Bottini\inst{\ref{iasf-mi}}
\and A.~Cappi\inst{\ref{oabo},\ref{nice}}
\and O.~Cucciati\inst{\ref{unibo},\ref{oabo}}           
\and I.~Davidzon\inst{\ref{lam},\ref{oabo}}   
\and P.~Franzetti\inst{\ref{iasf-mi}}   
\and A.~Fritz\inst{\ref{iasf-mi}}       
\and A.~Iovino\inst{\ref{brera}}
\and J.~Krywult\inst{\ref{kielce}}
\and V.~Le Brun\inst{\ref{lam}}
\and O.~Le F\`evre\inst{\ref{lam}}
\and D.~Maccagni\inst{\ref{iasf-mi}}
\and K.~Ma{\l}ek\inst{\ref{warsaw-nucl}},\inst{\ref{lam}}
\and F.~Marulli\inst{\ref{unibo},\ref{infn-bo},\ref{oabo}} 
\and M.~Polletta\inst{\ref{iasf-mi},\ref{marseille-uni},\ref{toulouse}}
\and A.~Pollo\inst{\ref{warsaw-nucl},\ref{krakow}}
\and L.A.M.~Tasca\inst{\ref{lam}}
\and R.~Tojeiro\inst{\ref{st-andrews}}  
\and D.~Vergani\inst{\ref{iasf-bo}}
\and A.~Zanichelli\inst{\ref{ira-bo}}
%
%
\and S.~Arnouts\inst{\ref{lam},\ref{cfht}} 
\and J.~Coupon\inst{\ref{geneva}}
\and G.~De Lucia\inst{\ref{oats}}
\and O.~Ilbert\inst{\ref{lam}}
\and T.~Moutard\inst{\ref{halifax},\ref{lam}}  
}

\institute{
INAF - Osservatorio Astronomico di Brera, Via Brera 28, 20122 Milano --  via E. Bianchi 46, 23807 Merate, Italy \label{brera}
\and Dipartimento di Scienza e Alta Tecnologia, Universit\`a degli studi dell'Insubria, Via Valleggio 11, I-22100 Como, Italy \label{insubria}
\and  Universit\`{a} degli Studi di Milano, via G. Celoria 16, 20133 Milano, Italy \label{unimi}
\and Aix Marseille Univ, Universit\'e Toulon, CNRS, CPT, Marseille, France \label{cpt}
\and Dipartimento di Matematica e Fisica, Universit\`{a} degli Studi Roma Tre, via della Vasca Navale 84, 00146 Roma, Italy\label{roma3} 
\and INFN, Sezione di Roma Tre, via della Vasca Navale 84, I-00146 Roma, Italy \label{infn-roma3}
\and INAF - Osservatorio Astronomico di Roma, via Frascati 33, I-00040 Monte Porzio Catone (RM), Italy \label{oa-roma}
\and Aix Marseille Univ, CNRS, LAM, Laboratoire d'Astrophysique de Marseille, Marseille, France  \label{lam}
\and Dipartimento di Fisica e Astronomia - Alma Mater Studiorum Universit\`{a} di Bologna, via Gobetti 93/2, I-40129 Bologna, Italy \label{unibo}
\and INFN, Sezione di Bologna, viale Berti Pichat 6/2, I-40127 Bologna, Italy \label{infn-bo}
\and INAF - Osservatorio Astronomico di Bologna, via Gobetti 93/3, I-40129, Bologna, Italy \label{oabo} 
\and Institute for Astronomy, University of Edinburgh, Royal Observatory, Blackford Hill, Edinburgh EH9 3HJ, UK \label{roe}
\and INAF - Istituto di Astrofisica Spaziale e Fisica Cosmica Milano, via Bassini 15, 20133 Milano, Italy \label{iasf-mi}
\and INAF - Osservatorio Astrofisico di Torino, 10025 Pino Torinese, Italy \label{oa-to}
\and Laboratoire Lagrange, UMR7293, Universit\'e de Nice Sophia Antipolis, CNRS, Observatoire de la C\^ote d’Azur, 06300 Nice, France \label{nice}
\and Institute of Physics, Jan Kochanowski University, ul. Swietokrzyska 15, 25-406 Kielce, Poland \label{kielce}
\and National Centre for Nuclear Research, ul. Hoza 69, 00-681 Warszawa, Poland \label{warsaw-nucl}
\and Aix-Marseille Université, Jardin du Pharo, 58 bd Charles Livon, F-13284 Marseille cedex 7, France \label{marseille-uni}
\and IRAP,  9 av. du colonel Roche, BP 44346, F-31028 Toulouse cedex 4, France \label{toulouse} 
\and Astronomical Observatory of the Jagiellonian University, Orla 171, 30-001 Cracow, Poland \label{krakow} 
\and School of Physics and Astronomy, University of St Andrews, St Andrews KY16 9SS, UK \label{st-andrews}
\and INAF - Istituto di Astrofisica Spaziale e Fisica Cosmica Bologna, via Gobetti 101, I-40129 Bologna, Italy \label{iasf-bo}
\and INAF - Istituto di Radioastronomia, via Gobetti 101, I-40129,Bologna, Italy \label{ira-bo}
\and Canada-France-Hawaii Telescope, 65--1238 Mamalahoa Highway, Kamuela, HI 96743, USA \label{cfht}
\and Department of Astronomy, University of Geneva, ch. d’Ecogia 16, 1290 Versoix, Switzerland \label{geneva}
\and INAF - Osservatorio Astronomico di Trieste, via G. B. Tiepolo 11, 34143 Trieste, Italy \label{oats}
\and Department of Astronomy \& Physics, Saint Mary's University, 923 Robie Street, Halifax, Nova Scotia, B3H 3C3, Canada \label{halifax}
}

\authorrunning{Mohammad et al.}

\offprints{\mbox{F.~G.~Mohammad},\\ \email{faizan.mohammad@brera.inaf.it}}

\abstract{We used the VIMOS Public Extragalactic Redshift Survey (VIPERS) final data release (PDR-2) to investigate the performance of colour-selected populations of galaxies as tracers of linear large-scale motions. We empirically selected volume-limited samples of blue and red galaxies as to minimise the systematic error on the estimate of the growth rate of structure $f\sigma_8$ from the anisotropy of the two-point correlation function. To this end, rather than rigidly splitting the sample into two colour classes we defined the red or blue fractional contribution of each object through a weight based on the $(U-V)$
colour distribution. 
Using mock surveys that are designed to reproduce the observed properties of VIPERS galaxies, we find the systematic error in recovering the fiducial value of $f\sigma_8$ to be minimised when using a volume-limited sample of luminous blue galaxies. We modelled non-linear corrections via the Scoccimarro extension of the Kaiser model (with updated fitting formulae for the velocity power spectra), finding systematic errors on $f\sigma_8$ of below 1-2\%, using scales as small as 5 \hmpc. We interpret this result as indicating that selection of luminous blue galaxies maximises the fraction that are central objects in their dark matter haloes; this in turn minimises the contribution to the measured $\xi(r_p,\pi)$ from the 1-halo term, which is dominated by non-linear motions. The gain is inferior if one uses the full magnitude-limited sample of blue objects, consistent with the presence of a significant fraction of blue, fainter satellites dominated by non-streaming, orbital velocities.  
We measured a value of $f\sigma_8=0.45 \pm 0.11$ over the single redshift range $0.6\le z\le 1.0$, corresponding to an effective redshift for the blue galaxies $\left<z\right>=0.85$. 
Including in the likelihood the potential extra information contained in the blue-red galaxy cross-correlation function does not lead to an appreciable improvement in the error bars, while it increases the systematic error.
}

\maketitle

\section{Introduction}		 									  

Over the past two decades, observations have established that the Universe is undergoing a period of accelerated expansion. The expansion history $H(z)$ is now well constrained by geometrical probes such as Type-1a supernovae \citep{riess98,perlmutter99}, Baryon Acoustic Oscillations \citep[BAO; e.g.][]{anderson14} in the clustering of galaxies and anisotropies in the Cosmic Microwave Background (CMB) \citep[e.g.][]{planck15}. In the framework of Einstein's General Relativity (GR), the observed $H(z)$ requires the inclusion of an extra contribution in the cosmic budget, in the form of a fluid with negative pressure, dubbed `dark energy'.  Current observations are compatible with the simplest form for this fluid, coinciding with Einstein's cosmological constant. Alternatively, however, one could also match the data by modifying the very nature of the gravitational equations. These two alternatives are degenerate when considering the expansion history of the Universe alone. Such a degeneracy can be lifted, in principle, by measurements of the growth rate of cosmological structure, which is sensitive to the gravity theory.

As the motions of galaxies respond to the gravitational potential, the velocity field can be used as a powerful probe of the growth of structure.  In galaxy redshift surveys, the line-of-sight velocity information becomes encoded in the redshift through the Doppler component which combines with the cosmological redshift, radially distorting galaxy positions in what is called `redshift space'. The amplitude of such `redshift-space distortions' \citep[RSD;][]{kaiser87}  can be quantified statistically by modelling their effect on two-point statistics. The linear component of the distortion is directly proportional to the linear growth rate of structure, $f(z)$, and motivates the interest in RSD as a powerful way to break the degeneracy between GR and alternative theories of gravity \citep[][]{guzzo08}.   

Measuring $f$ from RSD is however complicated by the non-linear component of the velocity field, which dominates on small scales ($<3 \mhmpc$) and is produced by high-velocity galaxies inside virialised structures, such as groups and clusters. This component has to be properly modelled if one wants to extract the linear growth rate signal, fully exploiting the data \citep[e.g.][]{reid14}.  To this end, early measurements used a modification of the original linear model for the redshift-space power spectrum derived by \citet{kaiser87}, empirically accounting for non-linear contributions through a Lorenzian (or exponential in configuration space) damping \citep[the `dispersion model',][]{peacock94}. Numerical tests have shown that for galaxy-sized haloes this model tends in general to deliver biased estimates of $f(z)$, up to $\sim10\%$ \citep{okumura11,bianchi12}.  This is clearly incompatible with the percent precision goals of modern redshift surveys, motivating extensive work on improved RSD models extending into the non-linear regime \citep[e.g.][]{scoccimarro04,taruya10,reid11,bianchi15,bianchi16,uhlemann15}. Some of these models have been applied to real data, with positive results \citep[e.g.][and references therein]{pezzotta16,delatorre16}.

Given the challenge of modelling the non-linear regime, we can attempt to reduce the importance of these regimes in the data.  One way to achieve this is through a linearisation of the density field by thresholding density peaks using the `clipping' technique.  We study this approach in a parallel work \citep[][in preparation]{wilson17}. 

Another way is to identify, if they exist, sub-populations of galaxies that by their very nature are less affected by non-linear motions. In \cite{mohammad16}, for example, we used numerical simulations to explore the use of galaxy groups and clusters as tracers of large-scale linear motions, modelling their redshift-space auto and cross-correlation functions. Although the group auto-correlation function yields the least biased results, it is penalised by the reduced statistics, due to the inevitably smaller number of galaxy groups that can be identified in a survey catalogue. The best compromise between statistical and systematic errors was obtained using the group-galaxy cross-correlation function, with systematic errors remaining smaller than 5\% also when including measurements down to 5 \hmpc.  The idea beyond these experiments is that of eliminating or reducing the weight of high-velocity galaxies in virialised structures in the computed two-point function. In the language of the halo model \citep{cooray02}, these are the objects defined as satellites, in contrast to central halo galaxies. In this picture, auto-correlating groups enhances the large-scale halo-halo term, minimising the contribution of 1-halo pairs affected by high relative velocities. 

Going beyond this would require identifying a sub-class of galaxies that are more numerous than groups, while still being objects that are central galaxies in a halo. One way to look for such a sample is to start from the observed colour dichotomy of galaxies. It is well known since almost forty years \citep[e.g.][]{dressler80} that  high density regions, where random high-velocity motions dominate, are preferentially inhabited by red galaxies. Conversely, blue galaxies tend to avoid these regions (at least out to $z\simeq 1$) and as such should be less affected by the non-streaming motions typical of groups and clusters.  Several works in the literature have compared the clustering of active (blue) and passive (red) galaxies at $z\simeq 0$ \citep{Madgwick03,zehavi05}. Quantitatively, halo occupation distribution (HOD) models indicate that red galaxies are likely to be satellites in massive dark matter haloes, while blue galaxies are typically central galaxies in haloes of lower mass \citep{guo14}. It is also well known that the segregation of galaxy colours and morphologies is mirrored by the kinematics of red and blue galaxies \citep{guzzo97}. In the local Universe, early-type (i.e. red S0s and elliptical) galaxies show larger random velocities and prominent `fingers-of-God' features than late-type (blue spiral and irregular) galaxies.  

In this paper we exploit this dichotomy using the newly released data of the VIMOS Public Extragalactic Redshift Survey (VIPERS) \citep{Guzzo14,Scodeggio16}.  Measurements of the growth rate from RSD out to $z=1$ using the VIPERS final data release have been presented recently using complementary techniques \citep{pezzotta16,delatorre16,hawken16}.  The selection of a catalogue of galaxy groups is under way (Iovino et al., in preparation) and is limited by the VIPERS angular mask. Given its broad selection function (essentially flux-limited), high sampling rate and extended photometric information, VIPERS is ideal to select sub-samples of galaxies based on properties such as  luminosity and colour. 

Here we perform joint analyses of the auto-correlation and cross-correlation statistics of the sub-samples.  The study is focussed on two populations selected by colour, red and blue, and further selected by luminosity to form volume-limited samples.

In Sect. 2 we describe the VIPERS dataset and the methodology used for sample selection. In Sect. 3 we describe the construction of the mock catalogues that we used for building covariance matrices and testing the accuracy of the estimators. The computation of the correlation function statistics is described in Sect. 4.  In Sect. 5 we present the redshift-space distortion models that we use to fit the data in Sects. 6 and 7.  The results and final conclusions are given in Sects. 8 and 9. Throughout the work we adopted the standard flat $\Lambda$CDM cosmological model with parameters $(\Omega_{\rm b},\Omega_{\rm m},h,n_{\rm s},\sigma_8)=(0.045,0.30,0.7,0.96,0.80)$.

\section{Observational data}	\label{sec:class}	 		  

\subsection{The VIPERS survey}									

\label{sec:data}

The VIPERS survey extends over an area of $ 23.5$ deg$^2$ within the W1 and W4 fields of the Canada-France-Hawaii Telescope Legacy Survey Wide (CFHTLS-Wide).  The VIMOS multi-object spectrograph \citep{lefevre03} was used to cover these two fields with a mosaic of 288 pointings, 192 in W1 and 96 in W4. Galaxies were selected from the CFHTLS-Wide catalogue to a faint limit of $i_{\rm AB}=22.5$, applying an additional $(r-i)$ vs $(u-g)$ colour pre-selection that efficiently
and robustly removes galaxies at $z<0.5$. Coupled with a highly optimised observing strategy \citep{scodeggio09}, this doubles the mean galaxy sampling efficiency in the redshift range of interest, compared to a purely magnitude-limited sample, bringing it to 47\%.

Spectra were collected at moderate resolution ($R\simeq 220$) using the LR Red grism, providing a wavelength coverage of 5500-9500$\smash{\mathrm{\AA}}$. The typical redshift error for the sample of reliable redshifts is $\sigma_z=0.00054(1+z)$, which corresponds to an error on a galaxy peculiar velocity at any redshift of 163~\kms. These and other details are given in the PDR-2 release paper \citep{Scodeggio16}. A discussion of the data reduction and management infrastructure was presented in \citet{garilli14}, while a complete description of the survey design and target selection was given in \citet{Guzzo14}.  The dataset used in this paper is an early version of the PDR-2 data, from which it differs by a few hundred redshifts revised during the very last period before the release. In total it includes $89\,022$ objects with measured redshifts. As in all statistical analyses of the VIPERS data, only measurements with quality flags 2 to 9 inclusive are used, corresponding to a sample with a redshift confirmation rate of $96.1\%$ \citep[for a description of the quality flag scheme, see][]{Scodeggio16}. In this work we used the absolute magnitudes derived for the VIPERS sample in \citet{davidzon16}, where spectral energy distributions (SED) were fitted to the extensive multi-band ancillary photometry available for the survey, as part of the VIPERS Multi-Lambda Survey \citep{moutard16a}.

\subsection{Colour classification}\label{sec:class_col}

To split the VIPERS sample into two blue and red sub-samples, we used the observed bimodal distribution of galaxy properties. \citet{haines16} give an extensive discussion of bimodality in the final VIPERS data as a function of spectral properties.  Here we used a criterion based on photometry, following \citet{fritz14}, where $UV=(M_U-M_V)$ colours and their dependence on redshift are described  \citep[see also][]{siudek17}. 

\begin{figure}
    \centering
		\includegraphics[scale=0.2]{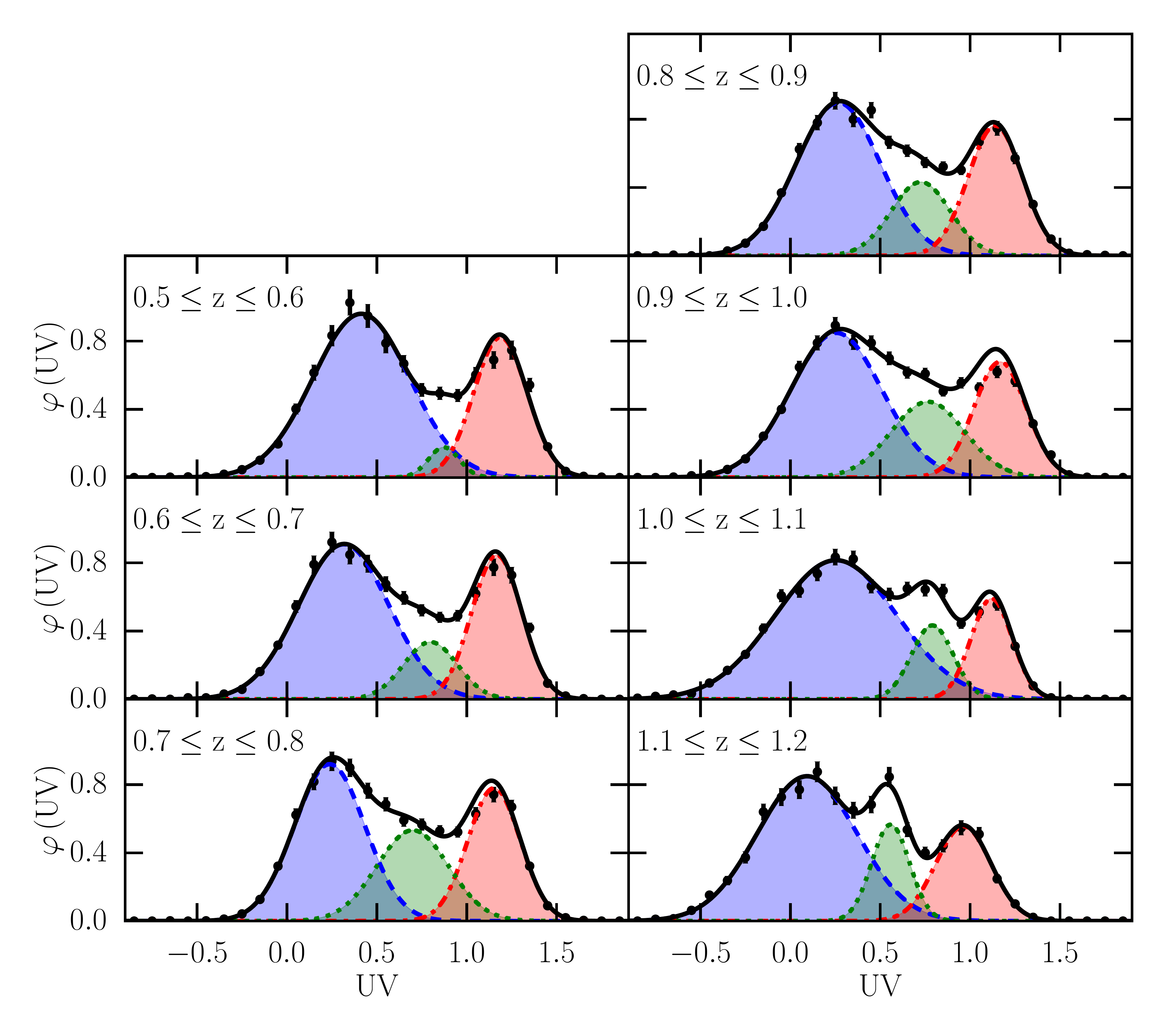}
		\caption{Normalised distribution of galaxy rest-frame $UV$ colour in VIPERS in the redshift range $0.5\le z\le 1.2$ (points). Statistical uncertainties include cosmic variance, estimated using the linear bias relation with $b=1.4$ (see Appendix \ref{app:err_pred}). The best-fit models (Eq. \ref{eq:3gauss}) are shown as black continuous lines. Contributions to the model from the galaxies belonging to the blue cloud, red sequence and green valley are plotted as blue dashed, red dash-dotted and green dotted lines, respectively.}\label{fig:bimod}
	\end{figure}

We modelled the $UV$ colour distribution with three Gaussian components. We note that the details of this split are not crucial for this work since our goal is essentially to separate a population dominating the high-density regions (the `red' galaxies) from the remaining class of bluer objects that should mostly populate the `field' and not to assess the reality of a third population with intermediate properties. Thus, the three-Gaussian fit simply characterises the three main populations of galaxies evident in the colour-magnitude diagram: the `red sequence', `green valley' and the `blue cloud'.  We performed the fit in redshift slices with width $\Delta z=0.1$ to account for redshift evolution. In each redshift bin, the measured $UV$ colour distribution was fitted with the three-Gaussian model 
	\begin{equation}
		\varphi\(UV,z\) = \varphi_b\(UV,z\)+\varphi_g\(UV,z\)+\varphi_r\(UV,z\)\; ,					\label{eq:3gauss}
	\end{equation}
where $\varphi_b\({UV},z\)$, $\varphi_g\({UV},z\)$ and $\varphi_r\({UV},z\)$ model the contribution to the overall colour distribution from the blue, green and red classes, respectively. Each term $\varphi_c\({UV},z\)$ on the right side of Eq. \eqref{eq:3gauss} was modelled as a Gaussian distribution,
	\begin{equation}
		\varphi_c\({UV},z\) = \frac{\mathrm{A}_c\(z\)}{\sqrt{2\pi}\sigma_c\(z\)}\exp{\[-\frac{\({UV}-\mu_c\(z\)\)^2}{2\sigma_c^2\(z\)}\]}\; .									\label{eq:gauss}
	\end{equation}
In Eq. \eqref{eq:gauss}, $\rm{A}_c$, $\mu_c$ and $\sigma_c$ are respectively the normalization factor, the mean and the dispersion of the Gaussian distribution. Figure \ref{fig:bimod} shows the histograms of the $UV$ colour distribution in different redshift bins along with the related best-fitting models. In computing the normalised distributions of the $UV$ colour, we weighted each galaxy to correct for the target sampling rate TSR and spectroscopic success rate SSR (both quantities are defined and discussed in details in Sect. \ref{incompleteness}).
We assigned a red and blue weight to each galaxy based upon the model fit to quantify the likelihood of being a member of the red or blue classes.  The weight is normalised such that $w_b + w_r = 1$, with
\begin{equation}
		w_{c}\(UV_\mathrm{g},z_{\mathrm{g}}\) = \frac{\varphi_{c}\(UV_\mathrm{g},z_i\)}{\varphi_{b}\(UV_\mathrm{g},z_i\)+\varphi_{r}\(UV_\mathrm{g},z_i\)}\; ,			\label{eq:col_weights}
	\end{equation}
where $UV_\mathrm{g}$ and $z_{\mathrm{g}}$ are the galaxy colour and redshift while $z_i$ is the corresponding redshift bin, that is $z_i-\Delta z/2\le z_{\mathrm{g}}\le z_i+\Delta z/2$ and the subscript -c denotes the blue or red colour type.

We stress here that in this analysis only two classes are considered.  Galaxies with intermediate `green' colours contribute to the red and blue samples with their corresponding weights. In practice, the result is similar to the usual binary blue-red classification based upon $UV$ colour. However, here green galaxies are not discarded but enter the measurements proportionally to their blue or red fractions. The advantage is twofold. We avoid introducing a sharp, arbitrary cut to separate red from blue objects and we keep all the objects of the catalogue. In this work, we weighted each galaxy by its corresponding colour weight $w_b$ or $w_r$ when computing statistics on the blue or red samples, respectively.

\begin{figure}
    \centering
		\includegraphics[scale=0.2]{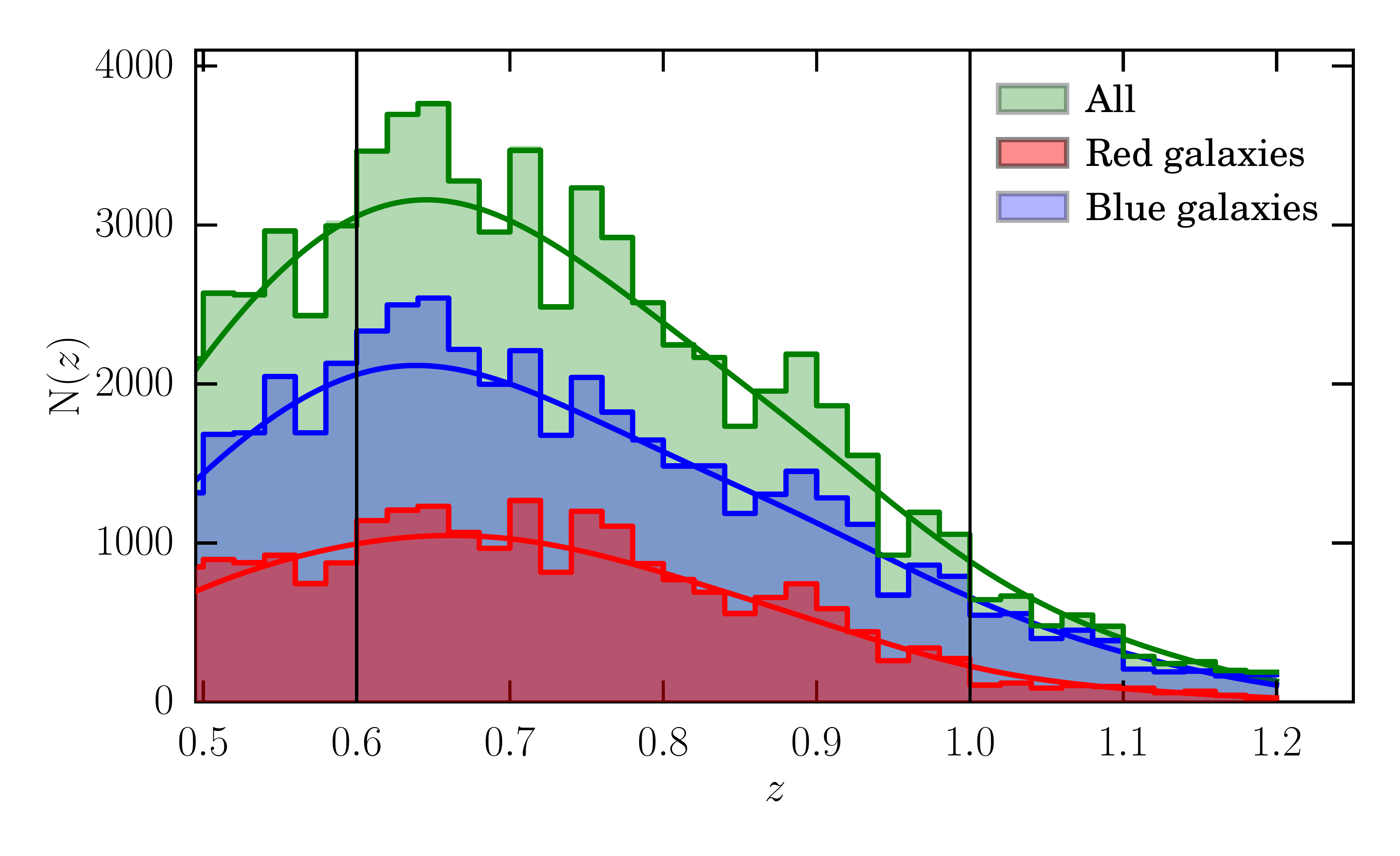}
		\caption{Un-normalised redshift distributions of VIPERS galaxies in the redshift range $0.5\le z\le 1.2$. The red and blue filled histograms show the observed number of blue and red galaxies respectively, i.e. when each galaxy is weighted by its blue $w_b$ or red $w_r$ colour weight only (see Eq. \ref{eq:col_weights}), resulting from our classification scheme (Sect. \ref{sec:class}). The distribution of all galaxies is also shown with green filled histogram. The continuous lines superposed on the histograms show the same distributions after convolving with a Gaussian kernel with $\sigma_z=0.07$. Vertical black  lines delimit the redshift range used in this analysis.}\label{fig:n-z_data}
	\end{figure}

The redshift distributions resulting from this classification are shown in Fig. \ref{fig:n-z_data} for the blue and red weighted samples along with the full sample of galaxies in VIPERS. The smoothed distribution using a Gaussian filter with width $\sigma_z=0.07$ are also shown in the same figure.
\subsection{Volume-limited samples}	\label{sec:vol-lim_data}

Selecting volume-limited samples from a flux-limited survey that covers an extended redshift range  entails making assumptions on how galaxies evolve within the redshift range. In the past this has been usually modelled through an empirical  luminosity evolution of the form
	\begin{equation}
		M\(z\) =  M_0 + M_1~z \; ,						\label{eq:lum-evol}
	\end{equation}
where $M_0$ is the absolute magnitude threshold one would assume at $z=0$, and, for the $B$ band and redshifts between 0 and $\sim 1$, a slope $M_1 \simeq -1$ was adopted to describe the average luminosity evolution of the full population of galaxies \citep[e.g. zCOSMOS:][]{lilly09}. This was empirically motivated by  the observed evolution of the characteristic luminosity $M^*$ in the same surveys, under the assumption of a pure luminosity evolution.  

Here we need to estimate the evolution parameters in Eq.~\ref{eq:lum-evol} for each of our colour-selected sub-classes. From now on, we restricted our analyses to $0.6\le z\le1.0$, a range which allowed us to build sufficiently large and fully complete volume-limited samples, given the VIPERS apparent magnitude limit. We worked under the same assumption that the comoving number density of galaxies in each class is preserved.
This is clearly not strictly true as over the restricted redshift range considered, $0.6\le z \le 1.0$, (a) we expect some objects to migrate from the blue cloud to the red sequence \citep[e.g.][]{gargiulo16,haines16,cucciati16} and (b) the merger rate is small but non-zero \citep[e.g.][]{fritz14}.  
In practice, these approximations have no impact on our conclusions, as our broad goal, as we shall show, is to maximize the fraction of central galaxies of galaxy-sized haloes.

Under these assumptions, we required the resulting comoving number density of galaxies in the selected samples to be constant with redshift, and computed the parameter values $(M_0,M_1)$ that give the corresponding integration limit of the luminosity function in Eq.~\ref{eq:lum-evol}.
We worked in bins of width $\Delta z = 0.05$, fixing the luminosity threshold ${M_B^\mathrm{th}(z)}$  to match the $90\%$ completeness value in the highest-redshift bin (i.e. $0.95\le z \le 1.0$) and computed the related reference comoving number density $n_{\mathrm{ref}}$. The luminosity threshold ${M_B^\mathrm{th}(z)}$ over the full range was then estimated as the one that keeps the comoving number density equal to this value: $n\(z\)=n_{\mathrm{ref}}$. We assumed ${M_B^\mathrm{th}(z)}$ to evolve linearly with redshift according to Eq. \eqref{eq:lum-evol}.

\begin{figure}
    	\centering
		\includegraphics[scale=0.2]{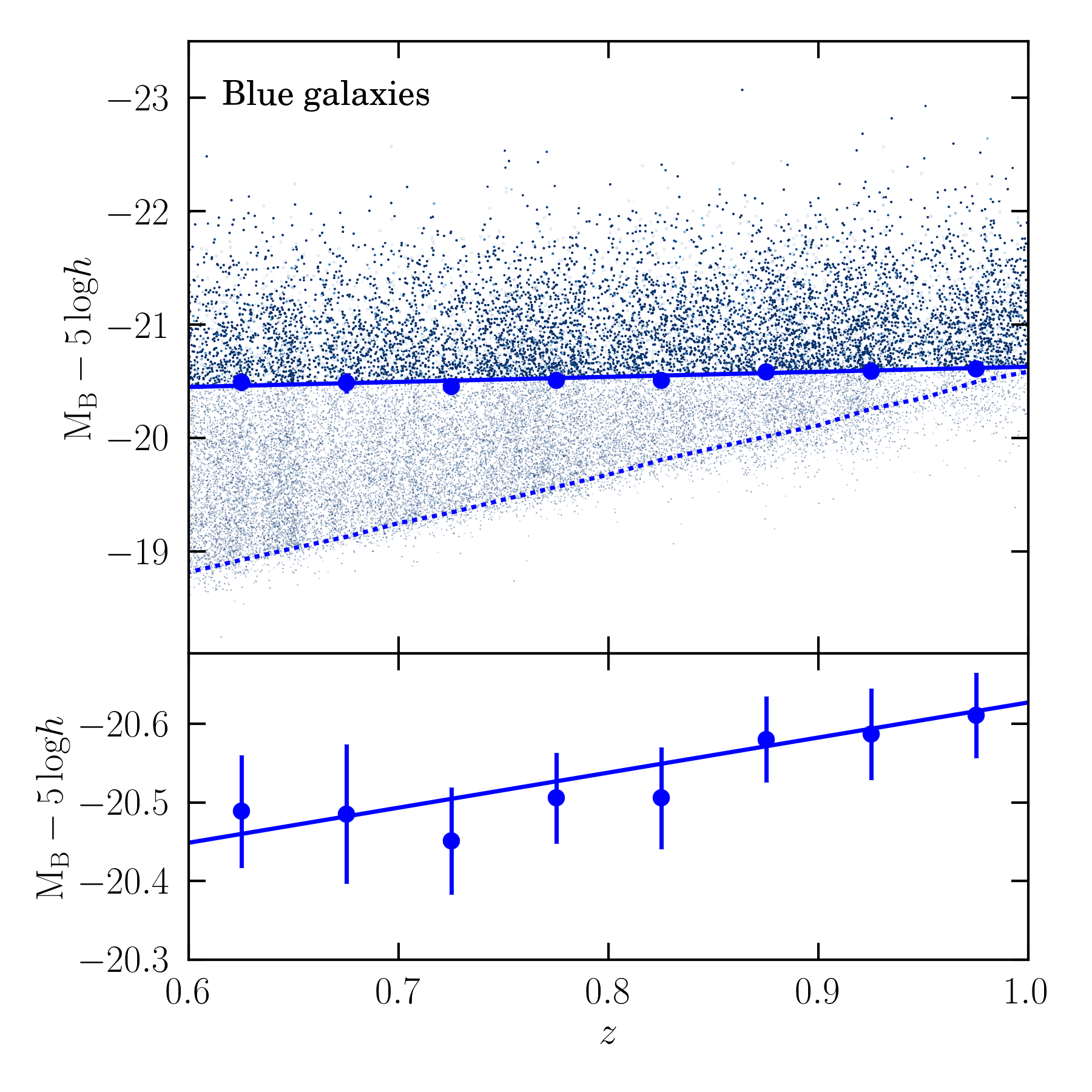}
		\caption{Magnitude-redshift diagram of VIPERS blue galaxies. Top panel: scatter plot in the magnitude ($M_B$)-redshift plane. Dark dots represent galaxies included in the volume-limited sample while the light dots show the ones excluded due to the luminosity threshold. Points show the luminosity threshold $M_B^{\mathrm{th}}(z)$ estimated by imposing a constant comoving number density as a function of redshift. The continuous lines are the best-fit model (Eq. \ref{eq:lum-evol}) to the points while the dashed curve represents the $90\%$ completeness level of the survey. Bottom panel: a zoom-in to highlight the agreement between data and model for the luminosity evolution.}\label{fig:vol-lim_blue_data}
	\end{figure}

\begin{figure}
    	\centering
		\includegraphics[scale=0.2]{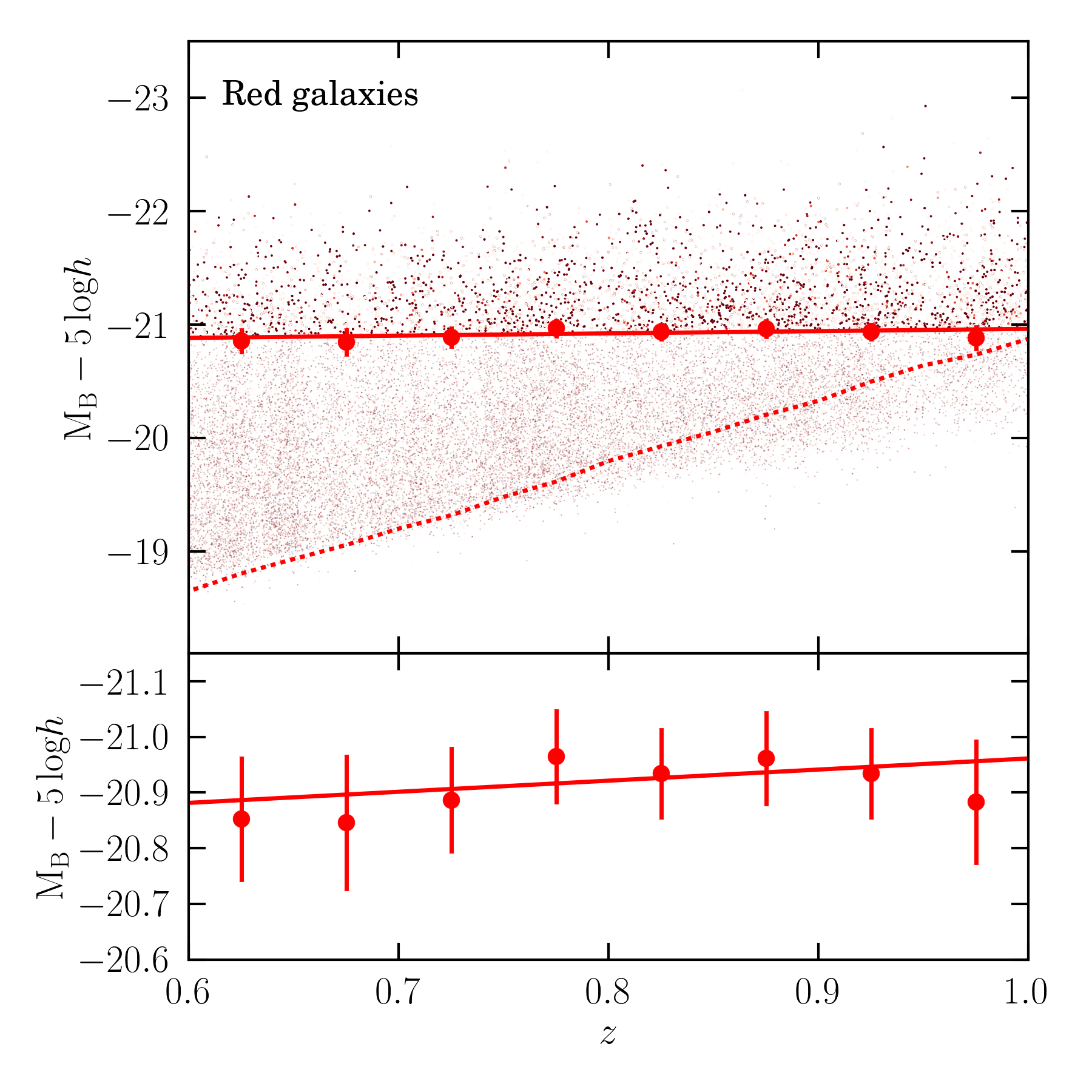}
		\caption{Same as in Fig. \ref{fig:vol-lim_blue_data} but here for red galaxies.}\label{fig:vol-lim_red_data}
	\end{figure}

The measured luminosity evolution function $M_B^{\mathrm{th}}\(z\)$ is shown in Fig. \ref{fig:vol-lim_blue_data} and in Fig. \ref{fig:vol-lim_red_data} for blue and red galaxy samples respectively, along with the related best-fit models.  The error budget $\sigma_n\(z\)$ on $n\(z\)$ takes contributions from the Poissonian shot-noise and the sample variance terms. The latter was estimated through linear theory predictions (see Appendix \ref{app:err_pred}) assuming a linear local and scale-independent bias $b=1.6$. With respect to the discussion in Sect. \ref{sec:class_col} we used a higher value here as the bias is known to be larger for more luminous galaxies \citep{marulli13,granett15,cappi15,diporto16}. The errors on the luminosity threshold $\smash{M_B^{\mathrm{th}}(z)}$ in each redshift bin were obtained by considering the values $\smash{M_B^{\mathrm{th}+}}$ and $\smash{M_B^{\mathrm{th}-}}$ corresponding to a comoving number density $n\(z\)+\sigma_n\(z\)$ and $n\(z\)-\sigma_n\(z\)$, respectively. The error on $\smash{M_B^{\mathrm{th}}(z)}$ was then obtained as $\smash{\sigma_{B}^{\mathrm{th}}(z)= (M_B^{\mathrm{th}+}-M_B^{\mathrm{th}-})/2}$. We finally fit the values of $\smash{M_B^{\mathrm{th}}\(z\)}$ inferred from the data with a linear model for the luminosity evolution in Eq. \eqref{eq:lum-evol}.

The best-fit evolution coefficients for the cases of red and blue galaxies, together with the main properties of the resulting volume-limited samples are listed in Table \ref{tab:lum-evol_data}. In this work we defined the effective redshift $z_{\rm{eff}}$ as the median of the distribution of the average redshift of all galaxy pairs with separations $3\mhmpc<s<50\mhmpc$.

\begin{table}
		\scriptsize
                \begin{center}
                        \begin{tabular}{	c						c					c				c					c						c							c}
                               \hline
                               \hline
						Type			&		$M_1$			&		$M_0$		&	$\rm{\chi/d.o.f}$		&	$N$		&	$N_{\mathrm{eff}}$	&		$z_{\mathrm{eff}}$\bigstrutup\bigstrutdown\\
                                \hline
						Red				&		$-0.20\pm0.14$		&	$-20.76\pm0.11$	&	$0.76$						&	$6,832$				&		$\sim3,652$		&		$0.84$\bigstrutup\\
						Blue			&		$-0.45\pm0.09$		&	$-20.18\pm0.07$	&	$0.77$						&	$14,673$			&		$\sim7,625$		&		$0.85$\\
                                \hline
                                \hline
                                \\
                        \end{tabular}
                \caption{Parameters characterising the volume-limited samples of red and blue galaxies in VIPERS within $0.6\le z\le1.0$. $\(M_0,M_1\)$ are the best-fit parameters for the luminosity evolution function in Eq. \eqref{eq:lum-evol} with corresponding reduced chi-square values $\rm{\chi/d.o.f}$. $N$ is the total number of galaxies included in the catalogue while $N_{\mathrm{eff}}$ is the effective number of galaxies, i.e. the sum of the related colour weights $w_b$ or $w_r$. Finally $z_{\mathrm{eff}}$ is the effective redshift of the sample.}\label{tab:lum-evol_data}
                \end{center}
        \end{table}

\section{VIPERS mock surveys}	 						
\label{sec:mocks}

We used a set of 153 independent VIPERS mock catalogues both to estimate the covariance matrix of clustering measurements and to test the impact of systematics arising from observational issues and RSD modelling. In our analysis we used two types of mock samples: 
\begin{enumerate}[i]
\item parent mocks - The light-cone galaxy catalogues with the VIPERS redshift distribution and rectangular sky coverage;
\item VIPERS-like mocks - The parent mocks with VIPERS survey geometry and application of the slit-assignment algorithm and redshift measurement error.
\end{enumerate}

\subsection{Mock construction}
The mocks were constructed from the Big MultiDark Planck \citep[BigMDPL,][]{prada12} dark matter N-body simulation using HOD prescriptions to populate  dark matter haloes with galaxies. The HOD parameters were calibrated using luminosity-dependent clustering measurements from the preliminary data release of VIPERS. The detailed procedure is described in \citet{delatorre13a,delatorre16}.  

The simulations were carried out in the flat $\Lambda$CDM cosmological model with parameters: 

$\(\Omega_m,\Omega_b,h,n_s,\sigma_8 \)=
\(0.307,0.048,0.678,0.96,0.823\)$. 
Since the resolution is not sufficient to match the typical halo masses probed by VIPERS, low mass haloes were added following the recipe proposed in \citet{delatorre13b}.
 
Central galaxies were placed at the halo centre with no peculiar velocities in the rest frame of the hosting halo.  Satellite galaxies were distributed within dark matter haloes according to an NFW profile \citep{Navarro97}. In addition to the hosting halo peculiar velocity, an additional random velocity component, drawn from a Gaussian distribution along each Cartesian direction, was assigned to the satellite galaxies. The velocity dispersion along each axis was computed following \citet{vandenbosch04} under the assumption of spherical symmetry and isotropy within dark matter haloes obeying an NFW density profile. This is clearly a delicate aspect in the case in which the mocks are used to test models of redshift-space distortions, as done for VIPERS, since the non-linear component of the velocity field is the most critical part of RSD modelling. We shall discuss this point further in this paper, when comparing results from the mocks and the real data.

Galaxy $B$-band luminosities and colours were assigned following the methods presented in \citet{skibba06} and \citet{skibba09}.  To summarise, halo occupation distribution model fits were carried out on the observed projected correlation functions measured in luminosity threshold samples, leading to an analytical luminosity- and redshift-dependent HOD parametrization \citep{delatorre13a}. The observed conditional colour bimodality $\langle UV | M_B\rangle$ in VIPERS was fitted with a double Gaussian distribution function. Using these fits, galaxies were placed in the simulation with the following recipe:
\begin{enumerate}
	\item For halo mass $m$ at redshift $z$, compute $\langle N_{\mathrm{cen}}(m | >M_{B,\mathrm{cut}}, z)\rangle$ and $\langle N_{\mathrm{sat}}(m | >M_{B,\mathrm{cut}},z)\rangle$, where $M_{B,\mathrm{cut}}$ is the absolute magnitude limit corresponding to $i=22.5$ at redshift $z$, and populate the given halo accordingly.
    \item Draw values of $M_{B}$ for the central and satellite galaxies by sampling from the cumulative distribution.  This is done by solving $\langle  N_{\mathrm{type}}(m | >M_{B},z)\rangle/\langle N_{\mathrm{type}}(m | >M_{B,\mathrm{cut}},z)\rangle=u$ for $M_{B}$, where $u$ is a uniform random number between 0 and 1 and the subscript `type' stands for `cen' or `sat' depending on the type of galaxy.    
    \item The rest-frame colour of the satellite and central galaxies is assigned with the relations \citep{skibba09} 
    \begin{equation} \label{eq:uvbsat}
    	\langle UV | M_B\rangle_{\mathrm{sat}} =  -0.19 M_B -2.25\; , 
    \end{equation}
and
	\begin{eqnarray}\nonumber
		&&\langle UV | M_B\rangle_{\mathrm{cen}} =  \langle UV | M_B\rangle_{\mathrm{all}}  \\ 
		&&+ \frac{n_{\mathrm{sat}}(M_B)}{n_{\mathrm{cen}}(M_B)}\left[\langle UV|M_B\rangle_{\mathrm{all}} - \langle UV | M_B\rangle_{\mathrm{sat}} \right]\; .
\end{eqnarray}
\end{enumerate}
Similarly to \citet{skibba09}, the coefficients in Eq. \ref{eq:uvbsat} have been set by trial and error, as to reproduce the observed segregation in the projected correlation function of red and blue galaxies.

\subsection{Volume-limited mock samples}		\label{sec:vol-lim_mocks}

Although the mock catalogues are found to be a good representation of the observed properties of the global galaxy population surveyed by VIPERS, they do not necessarily accurately reproduce the distributions of colour-selected galaxy samples.  We found that by following the procedure to construct volume limited samples described in Sect. \ref{sec:vol-lim_data}, we were unable to match both the number density and clustering amplitude of the blue and red samples in the mocks and data.  The main consequence of this mismatch is inaccuracy in the covariance matrices that we derive from the mocks.  As a compromise, we set the luminosity threshold to match the clustering amplitude.  This guarantees the accuracy of the cosmic variance contribution in covariance matrices.  This lead to a $\sim15\%$ deficit in the galaxy number density in the mocks with respect to the corresponding VIPERS samples.
However, we accounted for this mismatch of shot noise by modifying the covariance matrix (see Sect. \ref{sec:cov}).

To draw volume-limited mock samples from the flux-limited ones, we followed the same procedure adopted for real catalogues in Sect. \ref{sec:vol-lim_data}. We adopted a second-order polynomial in $z$ to better reproduce the mean luminosity evolution measured from our 153 mocks,
	\begin{equation}
		M\(z\) = M_0 + M_1z + M_2z^2\; .							\label{eq:lum-evol_mocks}
	\end{equation}
To match the clustering amplitude of red and blue galaxies in VIPERS we set the luminosity threshold for mock galaxies (both blue and red) to $M_B^{\mathrm{th}}=-20.50$ in the last redshift bin (See Fig. \ref{fig:best-fit_mps} where the measurements of the 2PCF multipoles of the luminous blue galaxies in VIPERS-like mocks and VIPERS data are plotted together).

\section{Two-point correlations}	\label{sec:meas}	  												
\begin{figure}
		\centering
		\includegraphics[scale=0.2]{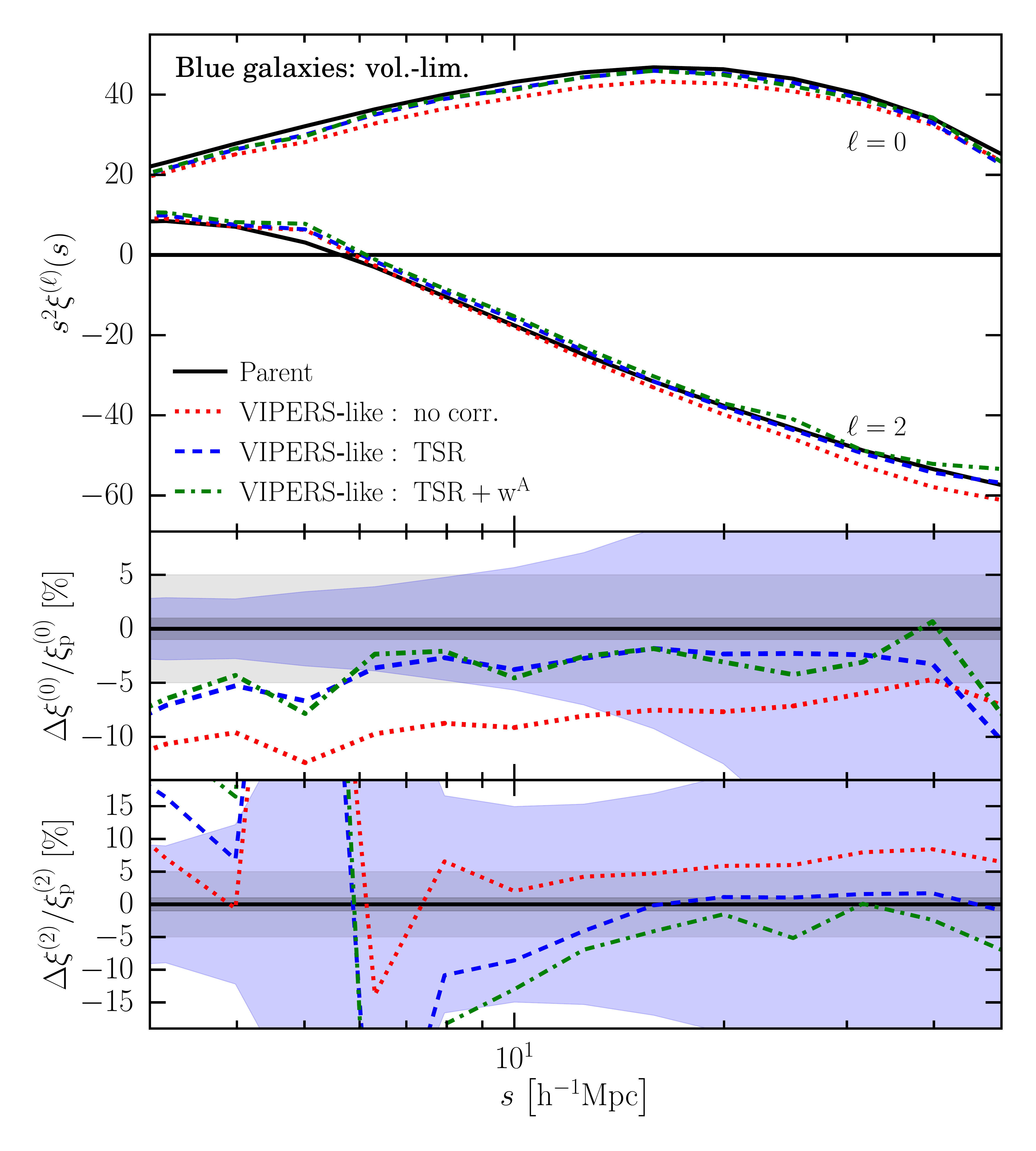}
		\caption{Impact of different corrections on the measured multipole moments of the two-point correlation function of blue galaxies in volume-limited mock samples. Top panel shows the measurements, while middle and bottom panels contain the relative systematic error on the monopole and quadrupole, respectively. Black continuous lines result from parent mocks. Red dotted lines are the raw estimates from VIPERS-like mocks. Blue dashed lines result from correcting the multipoles for TSR only; (also, no SSR correction is needed for the mocks). Green dash-dotted lines are the case when both TSR ($w_{\rm{TSR}}$) and angular ($w^{\rm{A}}$) weights are applied. Horizontal grey shaded bands in the middle and bottom panel delimit the $1\%$ and $5\%$ regions, while blue shaded regions show the 1-$\sigma$ error on the mean estimates of the multipole in parent mocks.}\label{fig:xi_blue_corr}
	\end{figure}   
	\begin{figure}
		\centering
		\includegraphics[scale=0.2]{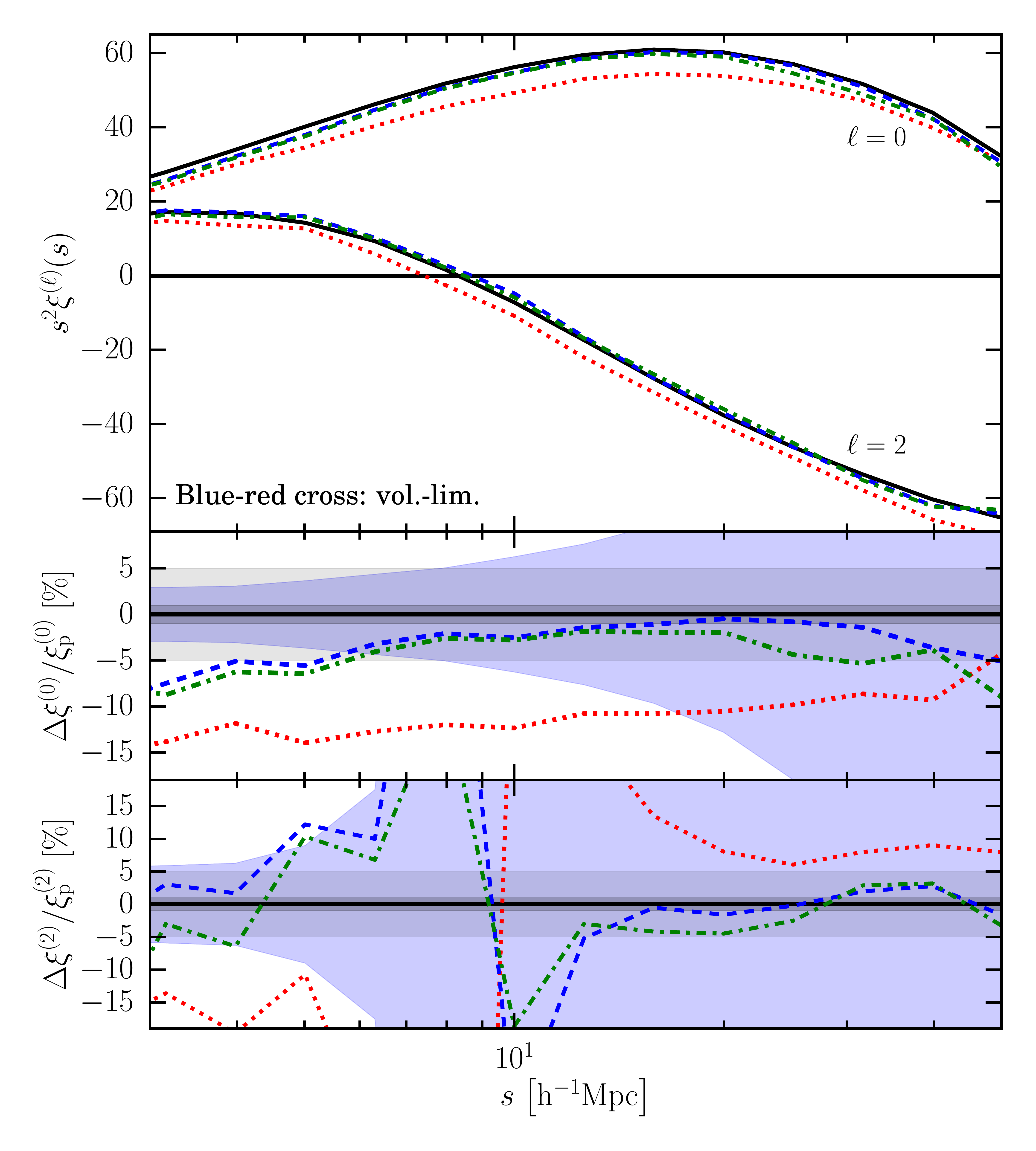}
		\caption{Same as in Fig. \ref{fig:xi_blue_corr} but for the multipole moments of the blue-red two-point cross-correlation function.}\label{fig:xi_cc_corr}
	\end{figure}
The anisotropic two-point correlation function was measured as a function of two variables, namely the angle-averaged pair separation $s$ and $\mu$, the cosine of the angle between the line of sight and the direction of pair separation. We describe here the estimator used to measure the anisotropic two-point correlation functions $\xi\(s,\mu\)$ and the method used to correct the measurements against different observational systematics.

\subsection{Estimator}
We used the minimum variance estimator proposed by \citet{landy93},
	\begin{equation}
		\xi\(s,\mu\) = \frac{\mathrm{DD}\(s,\mu\)-2~\mathrm{DR}\(s,\mu\)}{\mathrm{RR}\(s,\mu\)}+1\; .					\label{eq:LS}
	\end{equation}
In Eq. \eqref{eq:LS} $\mathrm{DD}$, $\mathrm{DR}$ and $\mathrm{RR}$ are respectively the data-data, data-random and random-random normalised pair counts. The random sample consists of points drawn uniformly from the survey volume characterised by the same radial and angular selection functions affecting the galaxy sample.

We obtained the cross-correlation function between volume-limited samples of blue and red galaxies by replacing the galaxy-galaxy pair count $DD$ with the blue-red galaxy cross-pair count $D_{\rm b}D_{\rm r}$ and the galaxy-random cross-pair count $2DR$ with $D_{\rm b}R+D_{\rm r}R$ with the subscript -b (-r) denoting the blue (red) galaxy type. The use of volume-limited samples allowed us to build a single random catalogue characterised by a comoving number density constant with redshift to probe the survey volume in virtue of the fact that both blue and red samples are affected by the same angular and radial selection functions.

In particular we used 200 linear bins in $\mu$ between $[0,1]$ with measurements sampled at the mid point of each bin in $\mu$. The pair separation $s$ was binned using logarithmic bins,
	\begin{equation}
		\log s_{i+1} = \log s_i + \Delta s_{\log}\; ,			\label{eq:log_bin}
	\end{equation}
with $\Delta s_{\log}=0.1$. The value of $s$ to which the measured correlation in each bin is referenced was defined using the logarithmic mean\footnote{In general, for a given logarithmic bin with edges $s_i$ and $s_{i+1}$ given by Eq. \eqref{eq:log_bin}, the logarithmic mean (Eq. \ref{eq:log_bin1}) is always smaller than the arithmetic mean and more closely matches the mean of the distribution of pairs.},
	\begin{equation}
		\log\langle s_i\rangle = \frac{\log s_i + \log s_{i+1}}{2}\; .																	\label{eq:log_bin1}
	\end{equation}

The measured anisotropic two-point correlation function $\xi^s\(s,\mu\)$ was then projected on the Legendre polynomials $L_{\ell}\(\mu\)$ in order to obtain the multipole moments of the two-point correlation function $\xi^{s,\(\ell\)}\(s\)$. However, given the discrete bins in $\mu$ we replaced the integral by the Riemann sum,
	\begin{equation}
		\xi^{s,\(\ell\)}\(s_i\) = \left(2\ell+1\right)\sum_{j=1}^{200}\xi^s(s_i,\mu_j)L_{\ell}(\mu_j)\Delta\mu\; .						\label{eq:meas_mps}
	\end{equation}
The number of bins in $\mu$ was deliberately taken high in order to have a good sensitivity to the direction of the pair separation, crucial for estimates of the quadrupole.

\subsection{Corrections for incompleteness}\label{incompleteness}

The target sampling rate (TSR) and spectroscopic success rate (SSR) result in incompleteness in the observed galaxy distribution with respect to the underlying one that systematically biases the two-point correlation function on large scales.  In particular, due to the slit placement constraints, the target sampling rate is lower in regions with a high density of galaxies on the sky.  This leads to a systematic reduction in the clustering amplitude.  The effect is even stronger for the more strongly clustered luminosity and colour sub-samples that we consider. Following the procedure presented in \citet{delatorre13a}, we corrected for the sampling effects by applying weights.  Each galaxy was weighted by the inverse of the effective sampling rate, $w_{\mathrm{ESR}}={\mathrm{TSR}}^{-1}\times {\mathrm{SSR}}^{-1}$ in addition to the colour weight corresponding to the blue or red sample selection,  $w_{b}$ or $w_r$.

The proper computation of the target sampling rate requires having the photometric parent sample. However, for sub-samples selected by luminosity, the parent sample is not known since it is defined using spectroscopic redshift.  Due to this limitation, we used the same target sampling rate estimated on the full flux-limited sample and used by \citet{pezzotta16} to analyse the full VIPERS sample.

The effective sampling rate affects the amplitude of the correlation function on large scales as shown in Fig. \ref{fig:xi_blue_corr} and Fig. \ref{fig:xi_cc_corr} for the multipoles of the auto correlation of blue galaxies and blue-red cross correlation respectively in the volume-limited mock samples. We do not show results for the auto-correlation of red galaxies, as the tests on the parent mocks will show that this class of galaxies produced very biased results (see Sect.~\ref{sec:tests}), and so will not be used to draw our final conclusions. However, we found that, in the case of volume-limited sample of red galaxies, the performance of the correction method was similar to the cases shown in Fig. \ref{fig:xi_blue_corr} and Fig. \ref{fig:xi_cc_corr}. The application of the weights corrected the monopole and quadrupole within $\sim5\text{-}7\%$ with the  exception of the zero-crossing region for the quadrupole.  

We found that the weights do not perform as well as shown in the full-sample analysis \citep{pezzotta16}.  This is due to the higher clustering of volume-limited samples, together with the fact that the weights are computed based upon the full sample.  The correction acts by upweighting galaxies according to the local projected density in the full sample, but since the sub-samples we considered are more clustered than in the full sample the weights do not fully account for the galaxies that were missed.

The pairs that are lost due to the slit placement constraints preferentially remove power on scales $<1\mhmpc$.   As described in \citet{delatorre13a} this bias may be corrected by the application of a weighting function to galaxy pairs that depends on angular separation (see Appendix \ref{ap:weights}).  We tested the application of angular weights using mock catalogues on the colour- and luminosity-selected samples and found no significant change in the measured monopole and quadrupole on the scales considered. We found that, as expected, the sparseness of our sub-samples amplifies the shot noise error and the uncertainties in the weight correction degrading the measurement. Therefore, for the subsequent analyses we did not apply the angular weights.

Taking into account the sampling rate corrections, the final pair counts are,
	\begin{subequations}
		\label{eq:pairs}
		\begin{align}
			DD\(s,\mu\)		&=	\sum_{i=1}^{N_g}\sum_{j=i+1}^{N_g}\rm{w_{c}^iw_{c}^jw_{ESR}^iw_{ESR}^j}\Theta_{ij}\(s,\mu\)\; ,		\\
			DR\(s,\mu\)		&=	\sum_{i=1}^{N_g}\sum_{j=1}^{N_R}\rm{w_{c}^iw_{TSR}^iw_{SSR}^i}\Theta_{ij}\(s,\mu\)\; ,								\\
			RR\(s,\mu\)		&=	\sum_{i=1}^{N_R}\sum_{j=i+1}^{N_R}\Theta_{ij}\(s,\mu\)\; .
		\end{align}
	\end{subequations}
In Eq. \eqref{eq:pairs} $\rm{w_{c}}$ is the galaxy colour weight related to the colour type c (blue or red) and $\Theta\(s,\mu\)$ is a step function equal to unity if $\log s\in[\log s_i-\Delta s_{\log}/2,\log s_i+\Delta s_{\log}/2]$ and $\mu\in[\mu_j-\Delta\mu/2,\mu_j+\Delta\mu/2]$ and zero otherwise.

In flux-limited galaxy samples, the radial selection function drops as one moves to higher redshifts. As a result the pair counts are dominated by the nearby galaxies with limited contribution from the more distant ones, even though the latter probe larger volumes. This motivates the use of $J_3$ weights \citep{hamilton93} in configuration space, or equivalently FKP weights \citep{feldman94} in Fourier space, to give an optimum balance between cosmic variance and shot noise in the two-point statistics. But the $J_3$ weighting scheme is found in practice to be ineffective for the flux-limited sample in VIPERS \citep{delatorre13a} and only makes the measurements noisier; we therefore did not include these weights in our measurements. In any case, the optimal weights are proportional to the inverse of the selection function except where shot noise dominates. Since we restricted the redshift range of our analysis to exclude the low-density tails, volume-limited samples should therefore give the main advantage claimed for optimal weighting, maximizing the effective volume and minimizing the sampling errors.

\begin{figure*}
		\centering
			\subfigure{\includegraphics[scale=0.37]{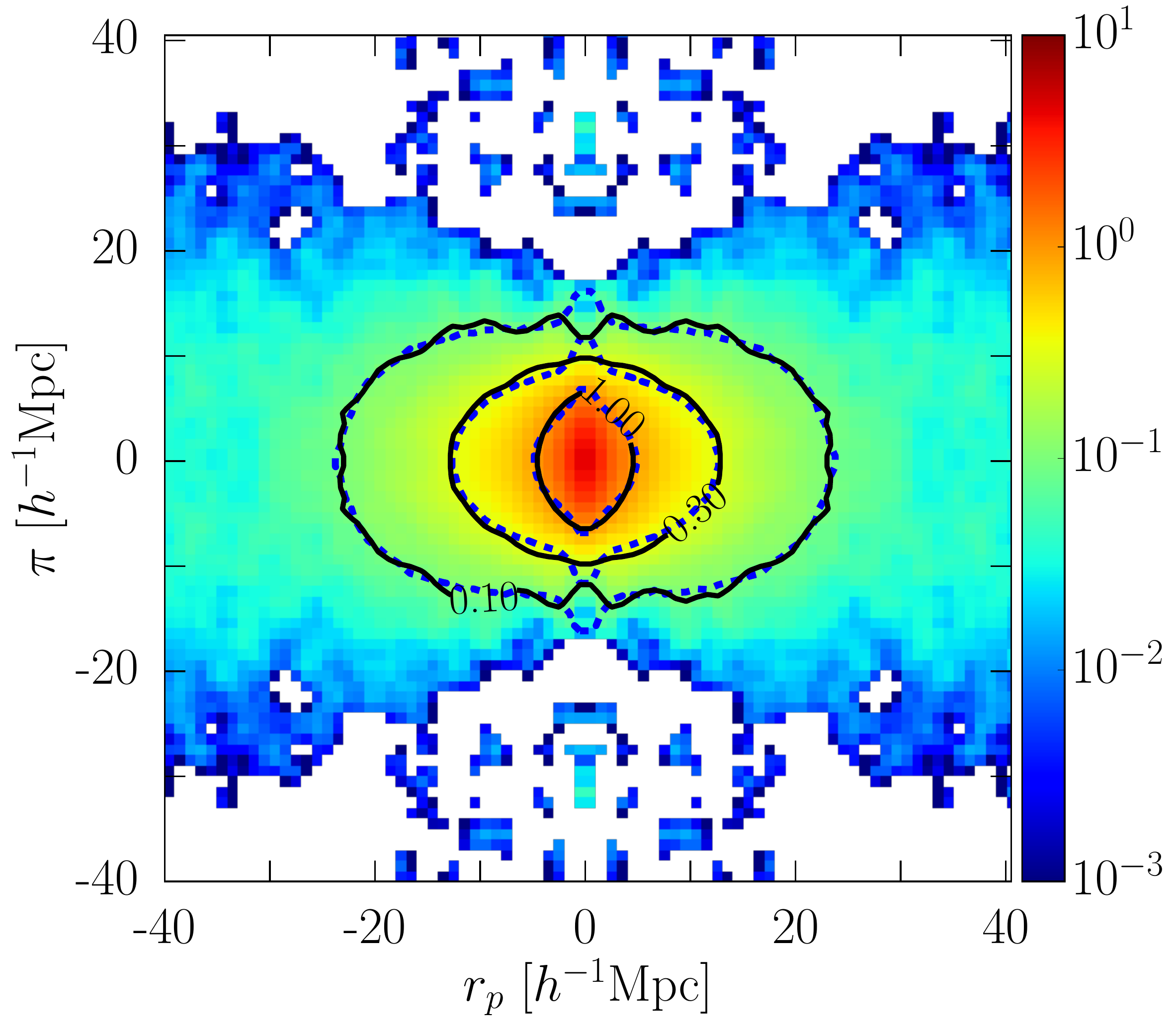}}
			\subfigure{\includegraphics[scale=0.37]{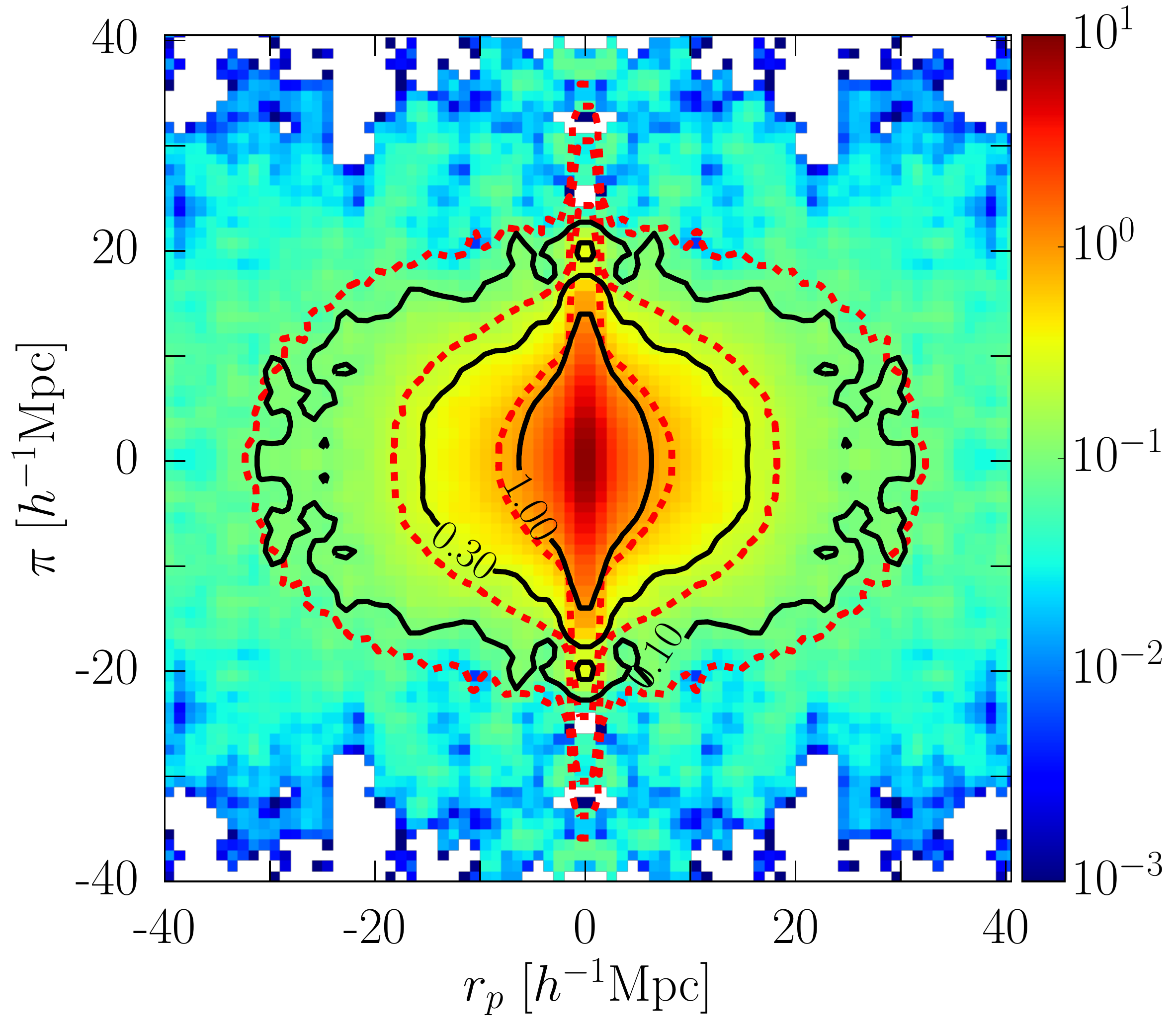}}
            
			\subfigure{\includegraphics[scale=0.37]{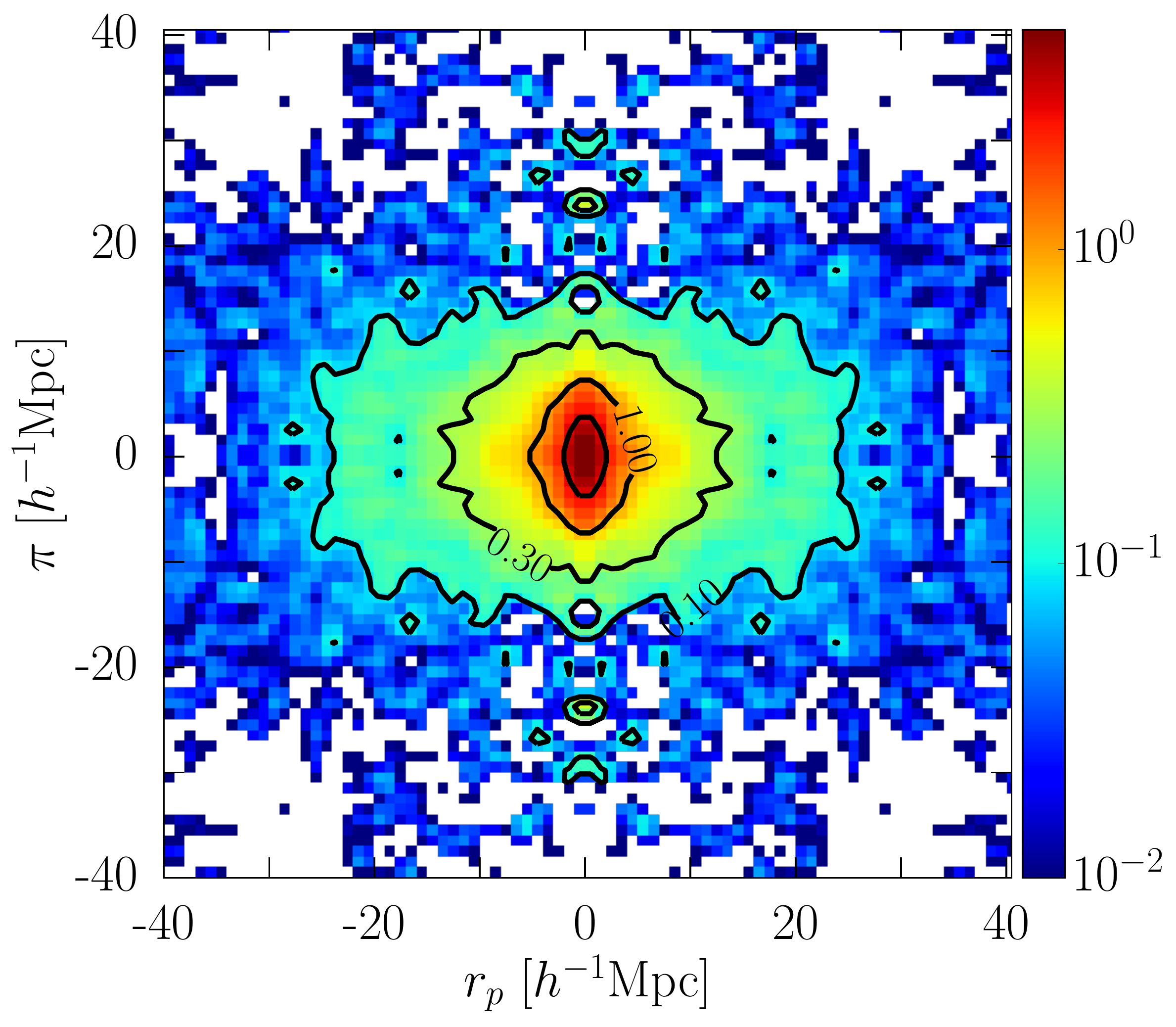}}
			\subfigure{\includegraphics[scale=0.37]{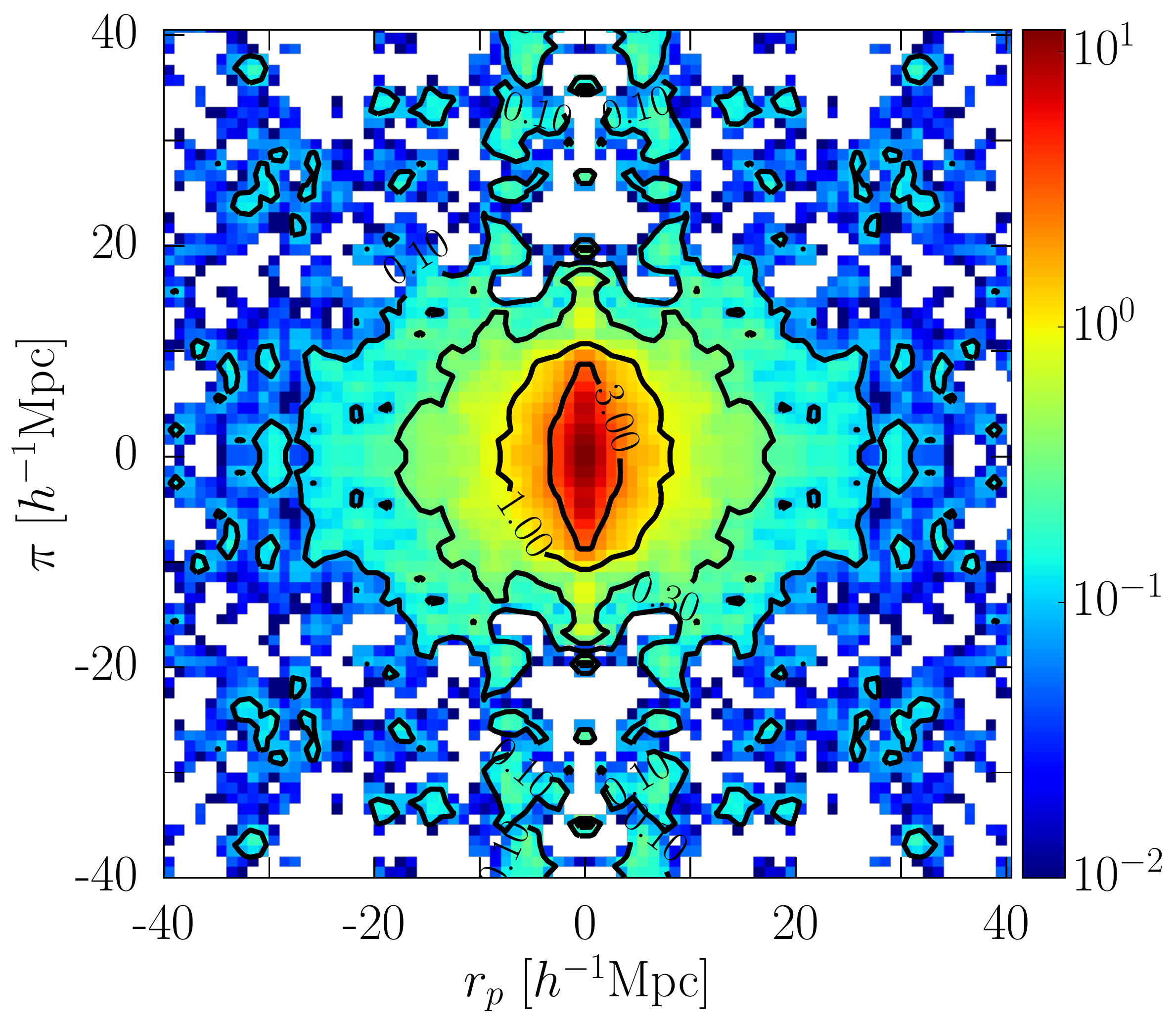}}
			\caption{\small{Redshift-space two-point correlation function $\xirppi$, measured at $0.6\le z\le 1.0$ from flux-limited (top row) and volume-limited samples (bottom row) of blue (left) and red (right) VIPERS galaxies (colour scale and solid contours). The measurements are binned in $1~\mhmpc$ bins in both directions and have been smoothed with a Gaussian filter with dispersion $\sigma=0.8~\mhmpc$. The more prominent small-scale stretching along the line of sight is clear in the clustering of red galaxies (right panels), which is almost absent for the blue galaxies (left panels). The dotted lines overplotted on the two top panels report instead for comparison the corresponding (un-smoothed) estimates from the mean of the 153 blue and red mock samples. The agreement of the blue mocks with the data is excellent.  Conversely, the red mocks show, in addition to their known slightly larger linear bias value, a significantly stronger small-scale stretching, indicating a higher non-linear velocity component with respect to the data (see text for discussion). In the two bottom panels the look-up table has been normalised as to get the same top colour at the peak value of $\xirppi$, while setting the bottom limit to $\xirppi=0.01$.  
}\label{fig:xirppi_vol-lim}}
        \end{figure*}

Figure \ref{fig:xirppi_vol-lim} shows the redshift-space two-point correlation functions $\xirppi$ for the blue and red populations both in the flux- and volume-limited sub-samples, computed applying the methodology and correction discussed above. The reduced small-scale FoG stretching for blue galaxies is evident. In the top panels we have also over-plotted the correlation function estimates obtained respectively from the mean of the blue and red mock samples. The agreement between data and mock samples for the blue population is remarkable on all scales.  This is not true for the red galaxies: first, the mock sample shows a higher amplitude, which was expected given its slightly higher linear bias. Secondly, the small-scale stretch of the contours produced by high-velocity dispersion pairs is significantly stronger.  We will have to keep this in mind when discussing the growth rate estimates based on the mock red galaxies; however, as we shall see, this difference will not change the main conclusions when comparing the blue and red samples.

\section{Theoretical models for RSD}	\label{sec:model}	 							

In large redshift surveys, observed redshifts result from a combination of the cosmological ones with the Doppler effect due to the line-of-sight component of the galaxy peculiar velocities. As a result the galaxy apparent positions $\mathbf{s}$ are distorted in the radial direction with respect to the real ones $\mathbf{r}$ if cosmological distances are inferred by means of observed redshifts,
	\begin{equation}
		\mathbf{s} = \mathbf{r}-f\(z\)\(\mathbf{u}\cdot\mathbf{e}_{||}\)\mathbf{e}_{||}\; .									\label{eq:model1}
	\end{equation}
In Eq. \eqref{eq:model1} $f(z)$ is the linear growth rate of structure, $\mathbf{e}_{||}$ is the unit vector along the line of sight and $\mathbf{u}$ is the scaled velocity field,
	\begin{equation}
		\mathbf{u} = -\frac{\(1+z\)}{f\(z\)H\(z\)}\mathbf{v}\; .																\label{eq:model2}
	\end{equation}
In terms of the overall matter density contrast $\delta=\rho/\bar{\rho}-1$ the mass conservation between true $\delta$ and redshift-space $\delta^s$ reads
 	\begin{equation}
 		\[1+\delta^s\]=\[1+\delta\]\left|\frac{d^3\mathbf{s}}{d^3\mathbf{r}}\right|^{-1}\; .									\label{eq:model3}
 	\end{equation}
In Eq. \eqref{eq:model3} $\left|d^3\mathbf{s}/d^3\mathbf{r}\right|$ is the Jacobian of the coordinate transformation in Eq. \eqref{eq:model1},
	\begin{equation}
		\left|\frac{d^3\mathbf{s}}{d^3\mathbf{r}}\right| = 1-f\(z\)\partial_{\|}u_{\|}\; .										\label{eq:model4}
	\end{equation}
Under the small-angle plane-parallel approximation in the regime where the density contrast and the velocity gradients are much smaller than unity, that is $\delta\ll 1$ and $\partial_\|u_\|\ll 1$ respectively, and the velocity field is irrotational, the mass conservation in Eq. \eqref{eq:model3} takes a much simpler form in Fourier space,
	\begin{equation}
		\delta^s\(\mathbf{k}\)=\delta\(\mathbf{k}\)+\nu^2f\(z\)\theta\(\mathbf{k}\)\; .										\label{eq:model5}
	\end{equation}
In Eq. \eqref{eq:model5} $\nu$ is the cosine of the angle between the wavevector $\mathbf{k}$ and the line of sight and $\theta=\nabla\cdot\mathbf{u}$ is the velocity divergence.

The galaxy bias $b$ is assumed here to be linear, local and scale-independent. Furthermore, the galaxy peculiar velocity field is assumed unbiased with respect to that of the overall matter. Eq. \eqref{eq:model5} can thus be written as
	\begin{equation}
		\delta_{\mathrm{i}}^s\(\mathbf{k}\) = b_{\mathrm{i}}\delta\(\mathbf{k}\)+\nu^2f\(z\)\theta\(\mathbf{k}\)\; ,			\label{eq:model6}
	\end{equation}
where the subscript -$i$ denotes the specific galaxy type considered. 

In the large-scale limit where $\delta=\theta$, Eq. \eqref{eq:model6} becomes
	\begin{equation}
		\delta_{\mathrm{i}}^s\(\mathbf{k}\) = \[b_{\mathrm{i}}+\nu^2f\(z\)\]\delta\(\mathbf{k}\)\; .							\label{eq:model7}
	\end{equation}
In the most general case of blue-red cross correlation explored in this work, the linear Kaiser model \citep{kaiser87} for the redshift-space cross power spectrum $P^s_{\mathrm{cr}}\(\mathbf{k}\)$ follows directly from Eq. \eqref{eq:model7},
	\begin{equation}
		P^s_{\mathrm{cr}}\(k,\nu\)=\[b_{\mathrm{b}}+\nu^2f\]\[b_{\mathrm{r}}+\nu^2f\]P_{\delta\delta}\(k\)\; .			\label{eq:model8}
	\end{equation}
In Eq. \eqref{eq:model8} $P_{\delta\delta}$ is the real-space matter power spectrum. The linear Kaiser model captures the enhancement in the galaxy clustering at very large scales. However it is not able to model the apparent clustering at smaller scales. \citet{peacock94} proposed the dispersion model, an empirical correction to the linear Kaiser model which accounts for the effect of fingers of God at small scales,
	\begin{equation}
		P^s_{\rm{cr}}\(k,\nu\)=D\(k\nu\sigma_{12}\)\times\[b_{\mathrm{b}}+\nu^2f\]\[b_{\mathrm{r}}+\nu^2f\]P_{\delta\delta}\(k\)\; .	\label{eq:model9}
	\end{equation}
The damping factor $D\(k\nu\sigma_{12}\)$ in Eq. \eqref{eq:model9} mimics the effect of pairwise velocity dispersion by suppressing the clustering power predicted by the linear Kaiser model. Here $\sigma_{12}$ is a scale-independent nuisance parameter which can be regarded as an effective pairwise velocity dispersion.

A more sophisticated model was derived by \citet{scoccimarro04} to extend the description of RSD at mildly non-linear regime treating separately the density and velocity divergence fields. In particular, dropping the approximation $\delta=\theta$, it follows from Eq. \eqref{eq:model6} that
	\begin{equation}
		P^s_{\rm{cr}}\(k,\nu\)=b_{\rm{b}}b_{\rm{r}}P_{\delta\delta}\(k\)+\nu^2\(b_{\rm{b}}+b_{\rm{r}}\)fP_{\delta\theta}\(k\)+\nu^4 f^2P_{\theta\theta}\(k\)\; ,					\label{eq:model10}
	\end{equation}
where $P_{\delta\theta}$ and $P_{\theta\theta}$ are respectively the density-velocity divergence and velocity divergence-velocity divergence power spectra. However, the Scoccimarro model in Eq. \eqref{eq:model10} fails in the description of small-scale motions within massive virialised structures dominated by high velocity galaxies with orbits that cross each other. This effect is included in the modelling in a similar way to the dispersion model by means of a damping factor,
	\begin{equation}
    	\begin{split}
		P^s_{\rm{cr}}\(k,\nu\) 	= D\(k\nu\sigma_{12}\)\times&\[b_{\rm{b}}b_{\rm{r}}P_{\delta\delta}\(k\)+\nu^2\(b_{\rm{b}}+b_{\rm{r}}\)fP_{\delta\theta}\(k\)+\right.\\
        &\left.+\nu^4 f^2P_{\theta\theta}\(k\)\]\; . \label{eq:model11}
		\end{split}
	\end{equation}
Although an improvement over the Kaiser model, the model in Eq. \eqref{eq:model10} is still an approximation. It is derived in the large-scale limit in the Gaussian case, while the probability distribution function (PDF) for the pairwise velocities is expected to be non-Gaussian at all scales. Furthermore, it neglects the scale dependence of the pairwise velocity PDF. Nevertheless, our tests in the following sections show that this model is able to capture the main effects in redshift space even at small scales provided that an appropriate galaxy sample, less affected by non-linear motions, is selected from the full galaxy population.

In \citet{pezzotta16}, where all galaxies are considered in the analysis, the modelling included the even more advanced extension represented by the so called Taruya or TNS model \citep{taruya10}, which takes into account the non-linear mode coupling between the density and velocity fields through additional corrections to the Scoccimarro model of Eq. \eqref{eq:model10} (which has the drawback of having extra degrees of freedom in the fit). Our goal here is complementary, that is to keep the modelling at a simpler possible level, but reducing the systematic biases through an optimised choice of galaxy tracers. For this reason we did not consider the TNS model.

The previous models also describe auto-correlation when $b_{\mathrm{b}}=b_{\mathrm{r}}=b$. RSD in the linear regime are degenerate with the linear growth rate $f$, the linear bias parameters $b$ and the amplitude of the linear matter power spectrum $\sigma_8$, so that one can constrain combinations of these parameters. Here we consider the combinations $f\sigma_8$ and $b\sigma_8$ once the input power spectra $P_{\delta\delta}$, $P_{\delta\theta}$ and $P_{\theta\theta}$ are renormalised by $\sigma_8^2$.

The input model for the linear matter power spectrum $P^{\mathrm{lin}}_{\delta\delta}$ was obtained using the `code for anisotropies in the microwave background' (CAMB) \citep{lewis00} which was then combined with the HALOFIT tool \citep{takahashi12} to predict the non-linear matter power spectrum $P_{\delta\delta}$ at the effective redshift of the galaxy sample. The density-velocity divergence cross power spectrum $P_{\delta\theta}$ and the velocity divergence-velocity divergence auto power spectrum $P_{\theta\theta}$ were obtained using the new fitting formulae calibrated on a large set of N-body simulations with various cosmologies \citep[DEMNUni:][]{carbone16} which are described in a companion work \citep[][in preparation]{bel17},
\begin{subequations}
                \label{eq:fitting}
                \begin{align}
                        \pdt            &=\[P_{\delta\delta}^{\rm{lin}}\(k\)P_{\delta\delta}\(k\)\exp\(-\frac{k}{k_{\delta\theta}^{\rm{cut}}}\)	\]^{1/2}\; ,\\
                        \ptt            &=\[P_{\delta\delta}^{\rm{lin}}\(k\)\exp\(-\frac{k}{k_{\theta\theta}^{\rm{cut}}}\)\]\; .
                \end{align}
        \end{subequations}
In Eq. \eqref{eq:fitting} $k_{\delta\theta}^{\rm{cut}}$ and $k_{\theta\theta}^{\rm{cut}}$ are defined as
	\begin{subequations}
		\label{eq:kcut}
        \begin{align}
			k_{\delta\theta}^{\rm{cut}}           &=\frac{1}{2.972}\sigma_8^{-2.034}\; ,\\
			k_{\theta\theta}^{\rm{cut}}           &=\frac{1}{1.906}\sigma_8^{-2.163}\; ,
		\end{align}
	\end{subequations}
with $\sigma_8$ being the amplitude of the linear matter power spectrum. These formulae are more general and represent an improvement over the previous expressions provided by \cite{jennings11}.

We adopted a Lorentzian functional form for the damping factor as it is found to provide the best description of N-body simulations and observations \citep{Zurek94,delatorre12,pezzotta16},
	\begin{equation}
		D\(k\nu\sigma_{12}\) = \frac{1}{1+\(k\nu\sigma_{12}\)^2}\; ,					\label{eq:model12}
	\end{equation}
with $\sigma_{12}$ being a free fitting parameter. The multipole moments of the anisotropic power spectrum $P\(k,\nu\)$ are given by
	\begin{equation}
		P^{\(\ell\)}\(k\) = \frac{2\ell+1}{2}\int_{-1}^{+1}P\(k,\nu\)L_{\ell}\(\nu\)d\nu\; .			\label{eq:rsd12}
	\end{equation}
In Eq. \eqref{eq:rsd12} $L_{\ell}$ is the Legendre polynomial of order $\ell$. The corresponding multipoles of the two-point correlation function can be easily obtained from $P^{\ell}\(k\)$ through
	\begin{equation}
		\xi^{\(\ell\)}\(s\) = i^\ell\int\frac{\dd k}{2\pi^2}k^2P^{\(\ell\)}\(k\)j_{\ell}\(ks\)\; ,																				\label{eq:rsd13}
	\end{equation}
with $j_\ell\(x\)$ being the spherical Bessel functions.

The dispersion and Scoccimarro models (Eq. \ref{eq:model9} and Eq. \ref{eq:model11} respectively) for the blue-red cross-correlation depend on four fitting parameters, namely $\{f\sigma_8,b_{\rm{b}}\sigma_8,b_{\rm{r}}\sigma_8,\sigma_{12}\}$. However, both models present a degeneracy between the three parameters $f\sigma_8$, $b_{\rm{b}}\sigma_8$ and $b_{\rm{r}}\sigma_8$. One way to break such a degeneracy is to estimate the relative bias $b_{\rm{12}}=b_{\rm{r}}/b_{\rm{b}}$ from the ratio of the projected two-point correlation functions at large  scales \citep{mohammad16}. But in our case the statistical errors on the measurements are sufficiently large that this approach does not give a stable estimate of $b_{\rm{12}}$. Alternatively, one can jointly fit the blue-red two-point cross-correlation function and one of the auto-correlation statistics. The price to pay is that the number of fitting parameters is increased to include the dispersion parameter $\sigma_{12}$ for the auto-correlation statistic $\{f\sigma_8,b_{\rm{b}}\sigma_8,b_{\rm{r}}\sigma_8,\sigma_{12}^{\rm{auto}},\sigma_{12}^{\rm{cross}}\}$. It is important to stress here that the previous considerations are valid only if redshift distributions of the blue and red samples have the same shape, resulting in the same effective redshift.

We fixed the redshift-distance relation to the fiducial model and neglected geometric distortions arising from the Alcock-Paczynski (AP) effect \citep{alcock79}. Including the AP effect would add two additional fitting parameters in the RSD model, the angular-diameter distance $D_A$ and the Hubble parameter $H(z)$, at the expenses of significantly larger statistical errors on the measurements of the cosmological parameters \citep[see e.g.][in preparation]{wilson17}. However, as shown in \citet{delatorre13a}, a change in the fiducial cosmology from WMAP9 to the Planck one results in a marginal variation in the estimates of $f\sigma_8$ of less than $1\%$,  small enough to be neglected in the total error budget of our analysis.

\section{Fitting method and data covariance matrix}		\label{sec:cov}		 
    \begin{figure}
    	\centering
		\includegraphics[scale=0.34]{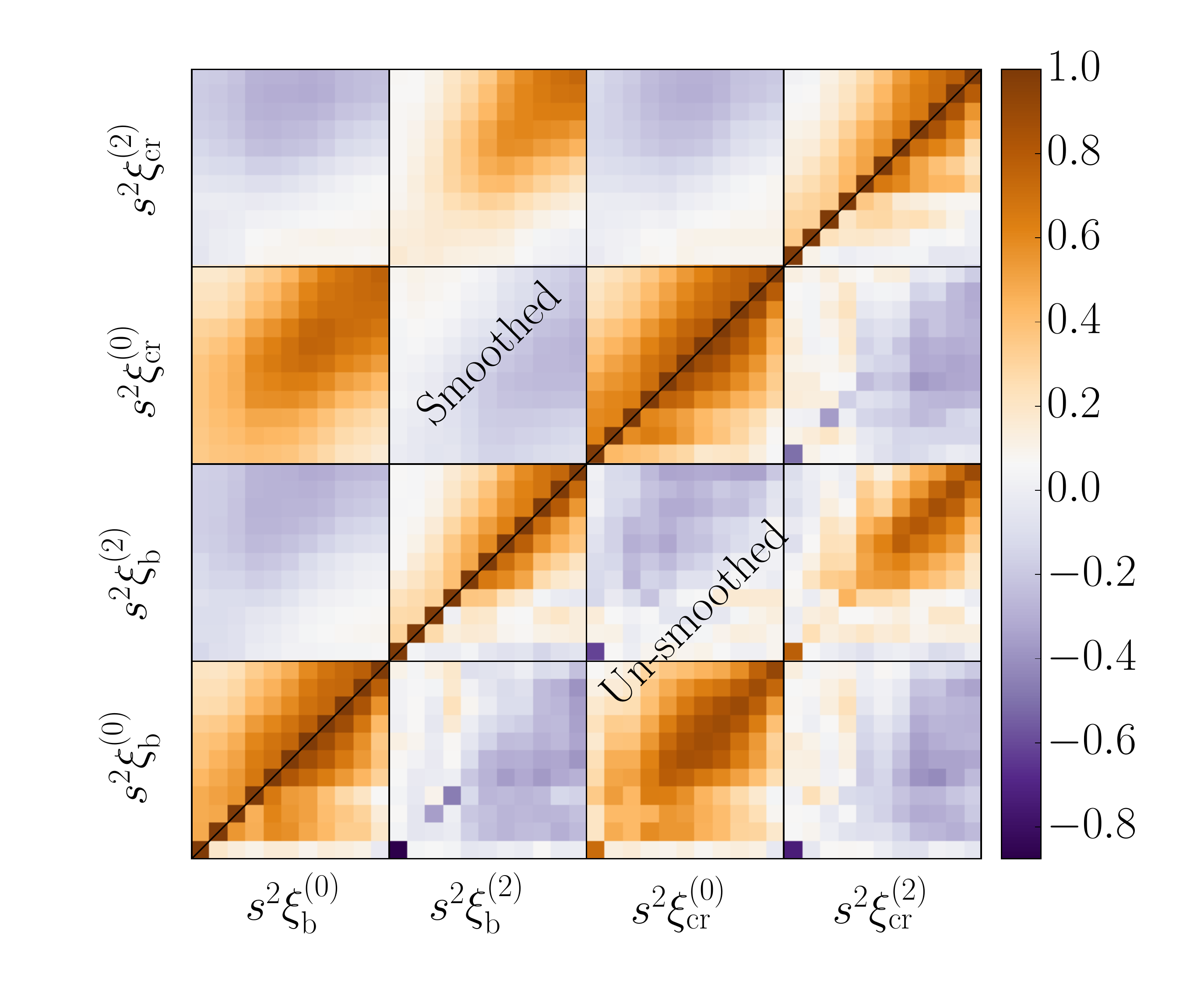}
		\caption{Joint data correlation matrix $\smash{ \rm R_{ij}=C_{ij}/\sqrt{C_{ii}C_{jj}} }$ between the first two multipoles of the auto-correlation of blue galaxies $\smash{ s^2\xi^{s,\(\ell\)}_{\rm{b}} }$ and their counterpart $\smash{ s^2\xi^{s,\(\ell\)}_{\rm{cr}} }$ for the blue-red cross correlation statistic in volume-limited samples. The pixels correspond to logarithmic bins spanning the scale range  $5-50\mhmpc$, the range used to obtain the final reference measurements of $f\sigma_8$ from the VIPERS data. The upper half has been box-car smoothed. }
        \label{fig:corr_ac+cc_data}
	\end{figure}
We fitted only the first two multipole moments of the redshift-space two-point correlation function, namely the monopole $\xi^{s,\(0\)}\(s\)$ and quadrupole $\xi^{s,\(2\)}\(s\)$, to estimate the linear growth rate of structure. However, rather than using the multipoles themselves, we considered the quantity
	\begin{equation}
		y^{s,\(\ell\)}\(s\) = s^2\xi^{s,\(\ell\)}\(s\)\; ,															\label{eq:fit0}
	\end{equation}
that has a smaller dynamical range and, therefore, easier to fit in the range of separations we consider here \citep{delatorre13a}.

\subsection{Data covariance matrix}

The clustering measurements in configuration space show a strong bin-to-bin correlation that needs to be taken into account when comparing data with theoretical models. This is quantified by means of the covariance matrix and its off-diagonal terms. To estimate the covariance matrix of the monopole and quadrupole moments of the auto- and cross-correlation functions we have used the 153 VIPERS mock catalogues. The covariance of a given quantity $y$ in two different measurement bins $i$ and $j$ is estimated as
	\begin{equation}
		C_{ij} = \frac{1}{N_s-1}\sum_{k=1}^{N_s}\[y_i^k-\langle y_i\rangle\]\[y_j^k-\langle y_j\rangle\]\; ,																\label{eq:cov}	
	\end{equation}
where $N_s$ is the number of mocks, $y_i^k$ is the measurement in bin $i$ drawn from the $k$-th mock while $\langle y_i\rangle$ denotes the mean of $y_i$ among its $N_s$ measurements. We fitted both the monopole and quadrupole of the two-point correlation function; thus, we also computed the cross-covariance of the multipoles.

The quantity of interest in our case is not the covariance matrix itself but rather its inverse matrix $C^{-1}$, in other words the precision matrix. Noise in the covariance matrix is amplified by the inversion process and leads to a biased estimate of the precision matrix \citep{hartlap07}. To account for this systematic error we followed \citet{percival14} and multiplied the generic element of the precision matrix by the factor $\(1-D\)$, with,
	\begin{equation}
		D = \frac{N_b+1}{N_s-1}\; ,																				\label{eq:cov1}
	\end{equation}
where $N_b$ is the number of measurement bins.  In the case of the correlation matrix shown in Fig. \ref{fig:corr_ac+cc_data}, $N_b=44$, $N_s=153$ and $1-D=0.71$.

The construction of the mock samples allowed us to match the clustering amplitude, but not the number density of VIPERS galaxies (see Sect. \ref{sec:vol-lim_mocks}).  This means that the shot noise contribution in the covariance is not accurate.  We modelled the covariance as the sum of two terms, the sample variance and shot noise,
    \begin{equation}
        C_{ij} = C_{ij}^{\rm{SN}} \delta_{ij} + C_{ij}^{\rm{SV}}\; .                                                                                               \label{eq:cov2}
    \end{equation}
The shot noise contribution was assumed to arise from a Poisson sampling process and is diagonal. We estimated this term using Monte Carlo realizations of a Poisson random field.  We generated a set of 153 un-clustered random samples containing a number of particles equal to the one in the galaxy catalogue under consideration.  Then using the monopole and quadrupole correlation functions measured in each Poisson random field we estimated the shot noise term of the covariance matrix.  The covariance matrix derived from the mocks was then modified by subtracting the shot noise term expected in the mocks and then adding the term corresponding to the number density of the VIPERS sample under consideration.

The estimated covariance matrix was rather noisy because of the limited number and the sparseness of the mock catalogues. To improve the estimation we used a box-car smoothing algorithm \citep{mandelbaum13} with a kernel of size $3\times3$ bins centred on the bin in consideration to smooth the off-diagonal terms of the covariance matrices.  Each sub-block of the covariance matrix was smoothed independently. This smoothing operation reduces the noise in the covariance matrix so the correction in Eq.~\eqref{eq:cov1} becomes only an approximation. In practice, the smoothed covariance matrix would be equivalent to using a larger number of mocks \citep{dodelson13}. Nevertheless, based on the tests shown in Appendix~\ref{app:smoothing}, we kept the correction factor as defined earlier.  This was a conservative choice as the correction acts to enlarge the error bars and Fig.~\ref{fig:smooth-raw_fs8} confirms that this procedure further stabilizes the systematic errors in our range of interest.  The condition number of the covariance matrix used for our reference estimate of the growth rate in Sect. \ref{sec:results} is $\sim10^{-3}$, well above the machine precision. Figure \ref{fig:corr_ac+cc_data} shows the correlation matrix $R_{ij}=C_{ij}/\sqrt{C_{ii}C_{jj}}$ before and after smoothing.

\subsection{Fitting method}

In Sect. \ref{sec:tests} we fit jointly the measured $y^{s,\(0\)}\(s\)$ and $y^{s,\(2\)}\(s\)$ [see Eq. \eqref{eq:fit0}] to estimate the parameters of both the dispersion and Scoccimarro models with a Monte Carlo Markov Chain (MCMC) algorithm.  The Markov chain explores the posterior distribution in the parameter space constrained by the data likelihood and parameter priors.  The data likelihood is given by
	\begin{equation}
		-2\ln\mathcal{L}\(\theta_p\) = \chi^2\(\theta_p\) = \sum_{i,j}\Delta_i\(\theta_p\)\ C_{ij}^{-1}\ \Delta_j\(\theta_p\)	\; ,		\label{eq:chi}
	\end{equation}
where $\chi^2$ is the goodness of fit, $\theta_p$ contains the set of model parameters and $\Delta_i$ is the difference between the measurement and the model predictions in bin $i$.  The data vector $y_i$ is a concatenation of the monopole $y^{s,\(0\)}$ and quadrupole $y^{s,\(2\)}$, restricted to the scales of interest. In particular, in Sect. \ref{sec:tests} we fit the measured multipoles between a varying minimum fitting scale $s_{\rm{min}}$ up to a maximum scale of $s_{\rm{max}}\simeq50\mhmpc$. Measurements on scales larger than $s_{\rm{max}}$ are affected by large statistical errors due to finite-volume effects. The performance of the RSD models was tested by gradually increasing the minimum fitting scale between $s_{\rm{min}}\simeq3\mhmpc$ up to $s_{\rm{min}}=10\mhmpc$. 

We adopted flat priors on each model parameter with bounds listed in Table \ref{tab:mcmc_priors}. As usual with an MCMC exploration of parameter space, marginalization over uninteresting degrees of freedom is achieved by ignoring those parameter values and simply dealing with the distribution of the MCMC samples over the parameter of interest -- namely $f\sigma_8$.

	\begin{table}
		\scriptsize
                \begin{center}
				\begin{tabular}{	c				c				c				c					c						c			c		}
                               \hline
                               \hline
										&		$f\sigma_8$	&	$\bb$	&		$\br$		&	$\sigma_{12}^{\rm{b}}$		&	$\sigma_{12}^{\rm{cr}}$		&	$\sigma_{12}^{\rm{r}}$	\bigstrutup\bigstrutdown	\\
                                \hline
						Min				&		$0.0$		&	$0.0$	&		$0.0$		&	$0.0$						&		$0.0$		&	$0.0$	\bigstrutup	\\
						Max				&		$2.0$		&	$3.0$	&		$3.0$		&	$5.0$						&		$5.0$		&	$10.0$\\
                                \hline
                                \hline
                                \\
                        \end{tabular}
                \caption{Lower (Min) and upper (Max) limits of the flat priors used to explore the fitting parameters space (see Sect. \ref{sec:model}) in Sects. \ref{sec:tests} and \ref{sec:results}.}\label{tab:mcmc_priors}
                \end{center}
        \end{table}

\section{Testing models on colour and luminosity mock VIPERS sub-samples}		\label{sec:tests}									

We used the VIPERS mock catalogues to test and optimise our RSD analysis applied to various galaxy sub-samples.

\subsection{Ideal case}					\label{sec:fit_ideal}
The first test used ideal mock VIPERS catalogues with no selection effects and observational errors to assess the importance of non-linear corrections. We refer to these ideal mocks as the VIPERS `parent' mock samples.

\subsubsection{Flux-limited samples}	\label{sec:fit_flux_ideal}
	\begin{figure}
    	\centering
		\includegraphics[scale=0.2]{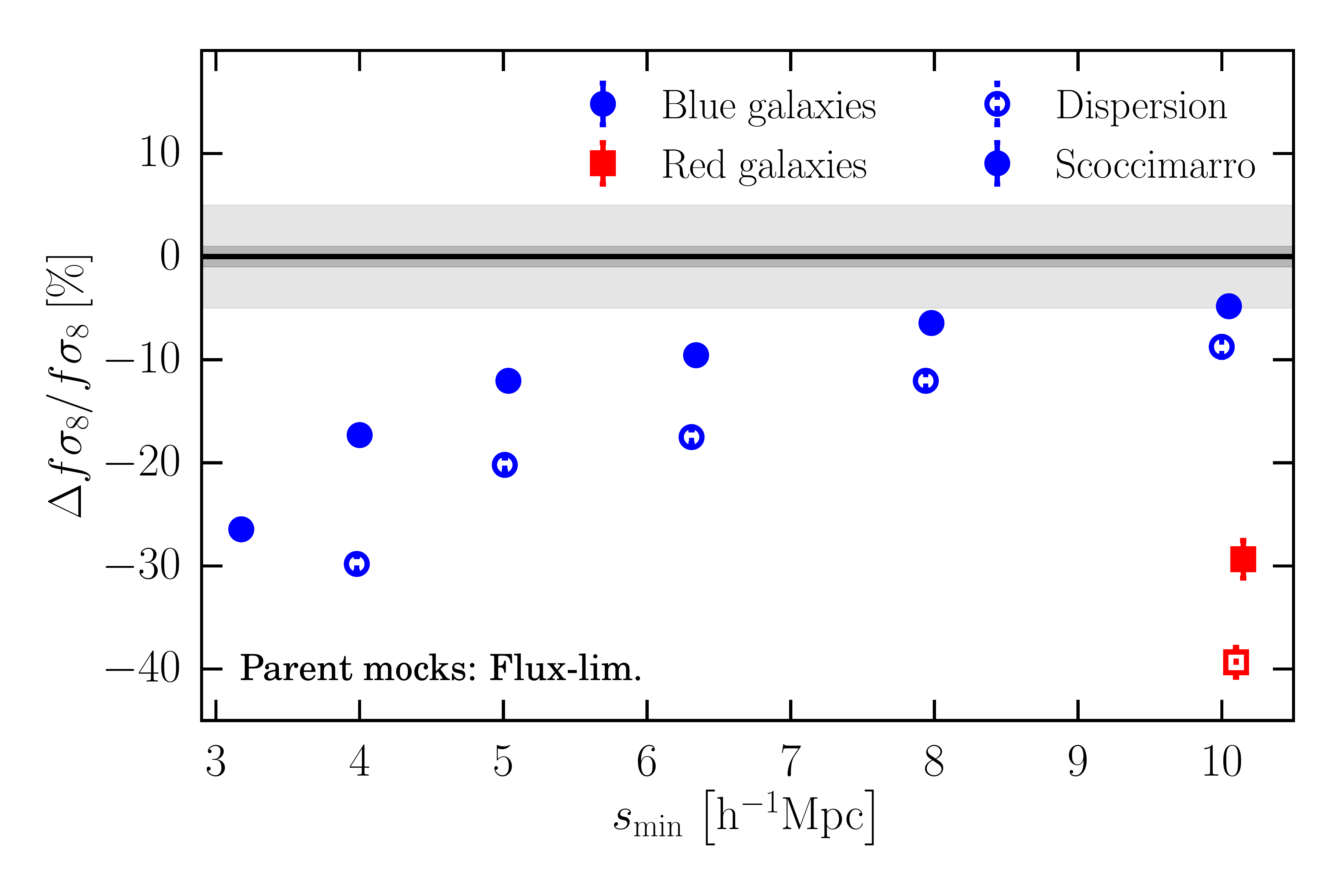}
		\caption{Systematic errors on the estimates of the growth rate $f\sigma_8$ when using the full parent mock flux-limited samples of blue and red galaxies in the redshift range $0.6\le z\le1.0$. The statistical errors, corresponding to the mean among 153 realizations, are also shown as vertical error bars although for the blue galaxies these are smaller than the size of points. The shaded regions correspond to $1\%$ and $5\%$ intervals on the growth rate, after marginalization over the hidden parameters of bias and velocity dispersion. We recall here that the apparent lack of red squares for $s_{\mathrm{min}}<10$ \hmpc is due to the large systematic errors from the red galaxies which are always greater than 45\%. Here, as in the following plots, points at the same $s_{\rm{min}}$ are slightly displaced for clarity.}\label{fig:flux-lim_syst_ideal_z_0.6-1.0}
	\end{figure}
Using the ensemble of flux-limited mock samples, we measured the monopole and quadrupole correlation function for red and blue samples.  We fixed the redshift range to $z=[0.6,1.0]$ that will also be used for the volume limited samples.  We fitted the measurements with the dispersion and Scoccimarro models and tested the dependence on the minimum scale used in the fit $s_{\mathrm{min}}$. The results are plotted in Fig. \ref{fig:flux-lim_syst_ideal_z_0.6-1.0}, in which we show the relative difference between the measured and expected values of $f\sigma_8$ as a function of $s_{\rm{min}}$. Since we are interested in estimating the systematic error, we consider the best estimate of $f\sigma_8$ that can be obtained from our 153 mocks and compare its deviations from the expected value. In principle, we should treat each mock as an independent realization to estimate the growth rate. The best estimate is then the mean of such 153 measurements of $f\sigma_8$. However, due to computational requirements we carried out a single fit on the mean correlation function multipoles of the mocks with appropriate covariance matrix.  As shown in Appendix \ref{app:fit}, both methods agree.

When using the red population, we measured a value of $f\sigma_8$ that is $\gtrsim30\%$ below the true one.  The blue galaxy sample performed better, but required the exclusion of scales below $s_{\rm{min}}=10\mhmpc$ in order to achieve a systematic error of $\sim5\%$ using the Scoccimarro model, which is consistent with previous works \citep[][]{delatorre12,pezzotta16}. The dispersion model is more biased than the Scoccimaro model at all scales.

For the flux-limited samples we did not considered combining these results with the blue-red cross-correlation since the blue and red populations are characterised by different effective redshifts. If we were to do so, the number of free parameters would increase to seven: $[f\sigma_8(z^{\mathrm{cr}}_{\mathrm{eff}}),b_{b}\sigma_8(z^{\mathrm{cr}}_{\mathrm{eff}}),b_{r}\sigma_8(z^{\mathrm{cr}}_{\mathrm{eff}}),\sigma_{12}^{\mathrm{cr}},f\sigma_8(z^{b}_{\mathrm{eff}}),b_{b}\sigma_8(z^{b}_{\mathrm{eff}}),\sigma_{12}^{b}]$  (see Sect.~\ref{sec:model}). This increase in the degrees of freedom fully erases the potential gain of the combination, which we shall explore only for the case of volume-limited samples.

\subsubsection{Volume-limited samples}\label{sec:fit_vol_ideal}
	\begin{figure}
    \centering
		\includegraphics[scale=0.2]{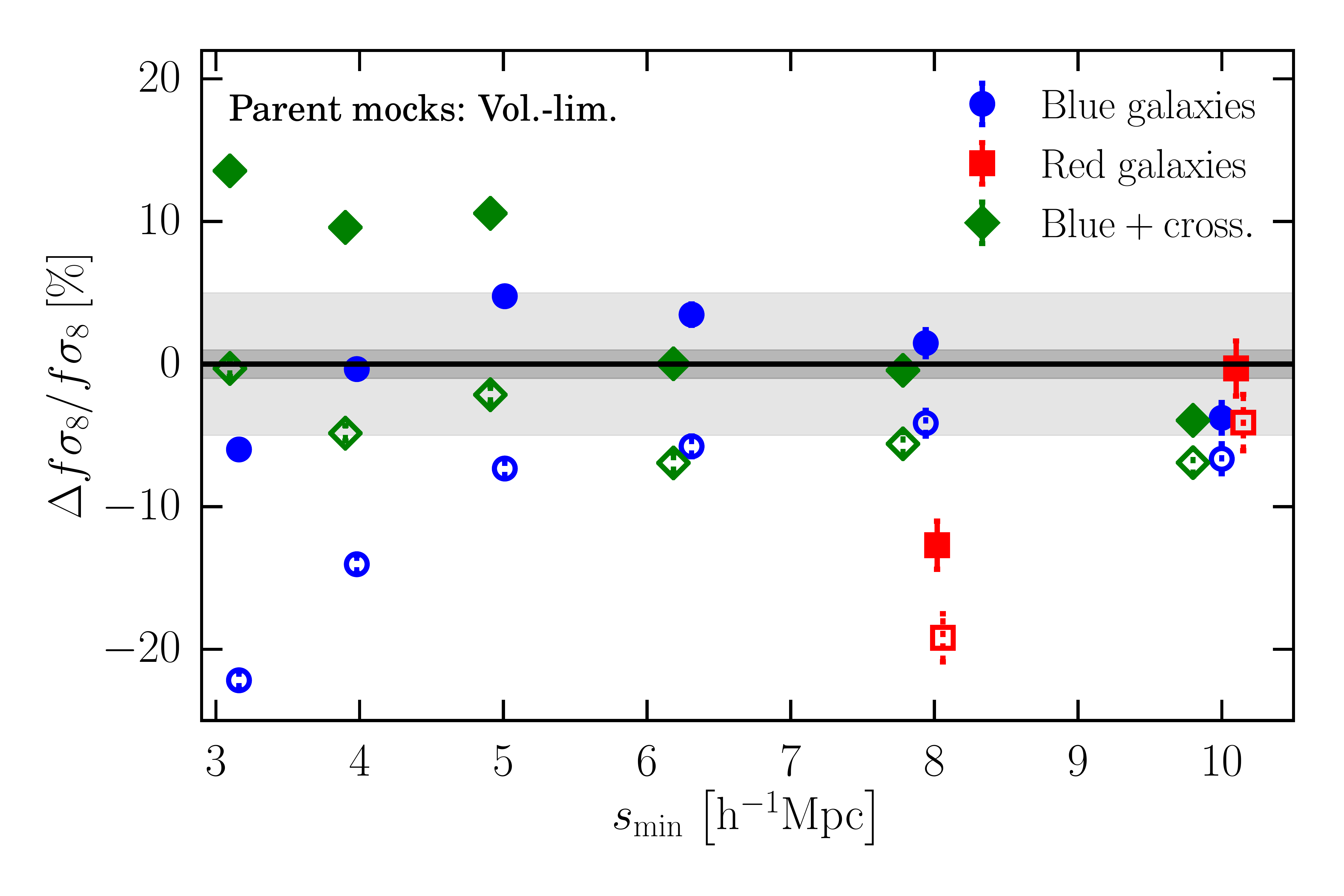}
		\caption{Same as in Fig. \ref{fig:flux-lim_syst_ideal_z_0.6-1.0} but now using volume-limited samples drawn from the parent mock catalogues. Empty and filled markers distinguish the dispersion and Scoccimarro models, respectively, as in Fig. \ref{fig:flux-lim_syst_ideal_z_0.6-1.0}. The green diamonds correspond to the joint fit of the blue galaxy auto-correlation and the cross-correlation of the two populations.\label{fig:vol-lim_syst_ideal}}
	\end{figure}
The tests using the flux-limited samples clearly suggest that we can reduce the weight of non-linearities by excluding red galaxies. Still, even when using blue galaxies alone a significant systematic under-estimate is evident, indicating a remaining non-negligible role of high-velocity-dispersion objects, that we interpret as the presence of `blue satellites' in dark matter haloes.  We then considered luminosity-selected blue and red samples to try to maximise the fraction of central galaxies within this population since the intrinsic luminosity cut excludes faint objects that are commonly satellite galaxies.

The corresponding results are plotted in Fig. \ref{fig:vol-lim_syst_ideal}, in the same form of the previous case. We also show results for the joint fit of the blue auto-correlation function with the blue-red cross-correlation. As before, red galaxies strongly underestimate the input growth rate parameter, although non-linear effects seem to be reduced when limiting the fit to scales larger than $s_{\rm{min}}=10\mhmpc$. The volume-limited sample of blue objects, instead, 
yielded systematic errors within $\pm 5\%$ down to the smallest explored scales, when the Scoccimarro correction (filled circles) was used. Also the simpler dispersion model delivered fairly good results down to $s_{\rm{min}}\simeq 6\mhmpc$.
The joint fit `blue + cross' also provided us with improved systematic errors (below $-5\%$ for $s_{\rm{min}}> 6\mhmpc$).
In all cases, as for the flux-limited samples presented in Sect. \ref{sec:fit_flux_ideal}, the relative difference between the dispersion and Scoccimarro models decreases considering higher values of $s_{\rm{min}}$.

\begin{figure}
	\centering
		\includegraphics[scale=0.2]{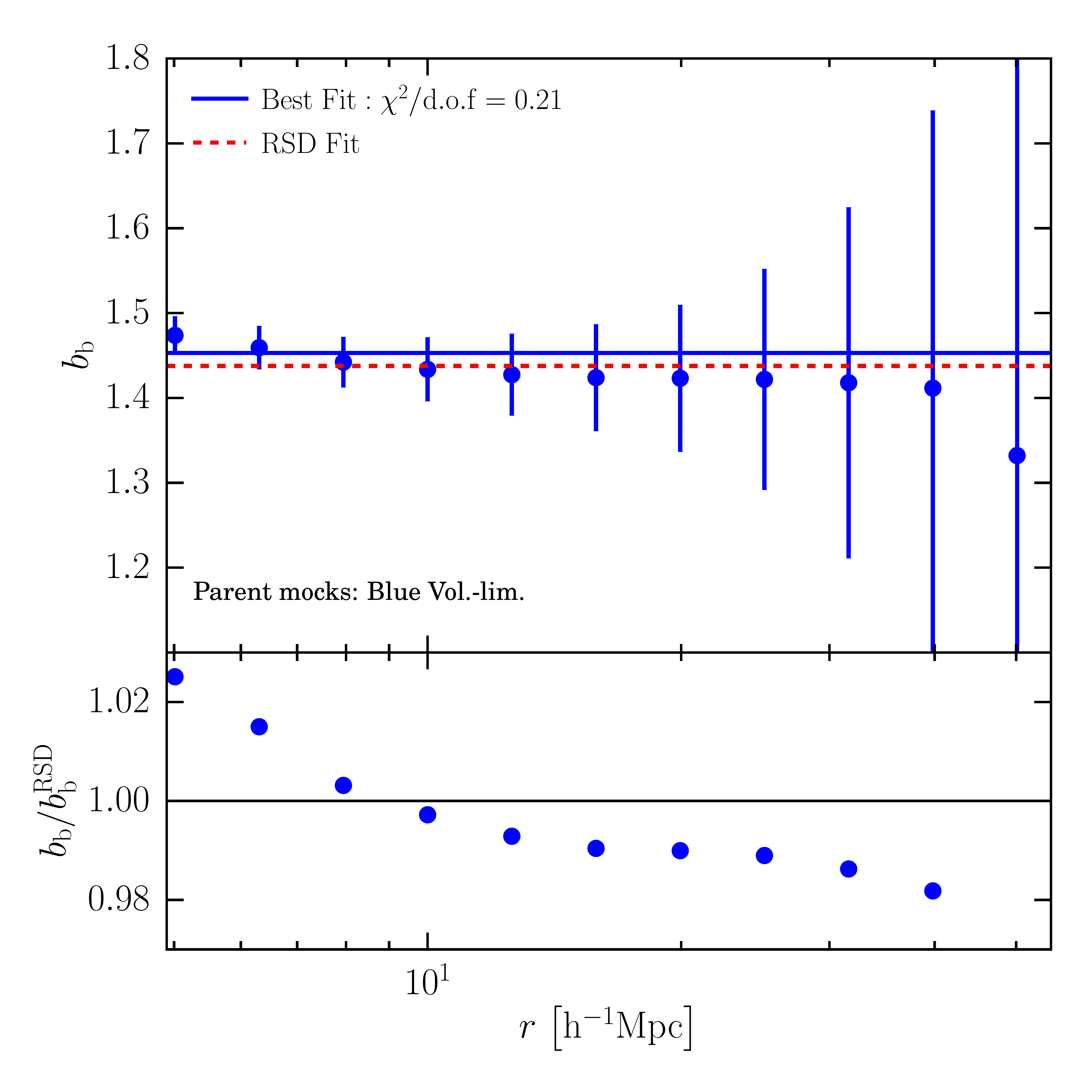}
		\caption{Linear galaxy bias of luminous blue galaxies in VIPERS parent mocks. Top panel: points with error bars show the mean measurement and the related 1-$\sigma$ error of the linear galaxy bias for the luminous blue galaxies in the 153 parent mocks, together with their best-fit with a constant value over $r=[5,50]\mhmpc$ (blue solid line). The red dashed line corresponds instead to the value yielded by MCMC using the Scoccimarro model, over the same rage of fitting scales. Bottom panel: the ratio of the measurements and the red dashed line in the top panel.}\label{fig:bias_blue}
\end{figure}
Our modelling assumes that the galaxy distribution traces the overall matter density field through a local, linear and scale-independent bias parameter $b$. One may wonder whether the excellent performance of the adopted RSD model for luminous blue galaxies could be due to an accidental cancellation of systematics from an inadequate dynamical model and a simplistic bias model. In the mock catalogues we can measure the galaxy bias as a function of scale,
    \begin{equation}
        b\(r\)= \left[\frac{\xi_g\(r\)}{\xi_m\(r\)}\right]^{1/2},                                   \label{eq:bias}
    \end{equation}
where $\xi_g(r)$ is the galaxy real-space correlation function and $\xi_m(r)$ is the non-linear matter correlation function used in the RSD model (see Sect.~\ref{sec:model}). We measured the bias of the luminous blue galaxy sample using the parent mock catalogues.  The mean measurement of $b(r)$ in the mocks is plotted in Fig. \ref{fig:bias_blue}.  The best-fitting bias inferred from the RSD analysis using scales down to $5\mhmpc$ is over-plotted.  It is remarkable that the inferred bias matches the real space measurement within $\sim2\%$.  This agreement gives us confidence that the local and scale independent bias assumption is justified and does not introduce a significant systematic error in the RSD analysis.

\subsection{Understanding the performances of volume-limited samples}
    \begin{figure}
    \centering
    	\includegraphics[scale=0.18]{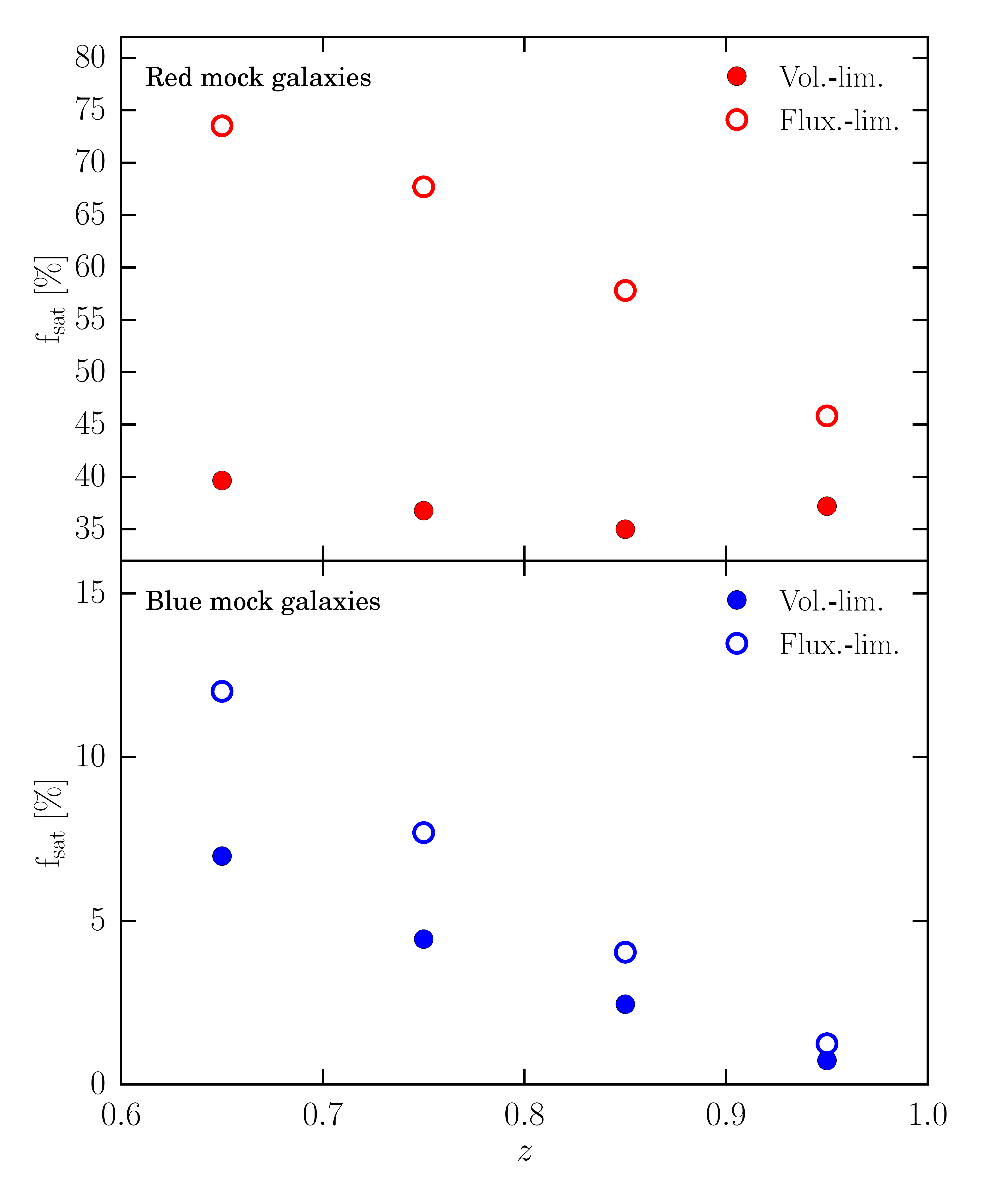}
    	\caption{Satellite fraction of colour selected blue and red galaxies in VIEPRS-like mocks. The halo occupation distribution (HOD) model used to build the mock catalogues identifies galaxies as either satellite or central. Plotted are the fraction of satellite galaxies $\mathrm{f_{sat}}$ in the flux- (empty markers) and volume-limited (filled markers) VIPERS-like mock samples for red (top panel) and blue (bottom panel) galaxies. In both panels $\mathrm{f_{sat}}$ is measured in redshift bins of width $\Delta z=0.1$ between $0.6\le z\le1.0$.}\label{fig:sat_frac_mocks}
    \end{figure}
These results clearly show that luminous blue galaxies preferentially trace large-scale, coherent, linear flows with little velocity dispersion, yielding the least biased estimates of $f\sigma_8$ that we could obtain.  This is consistent with our conjecture that the colour and luminosity selections we have applied mainly select galaxies that are likely to be central objects of their dark matter haloes in the halo occupation distribution picture.  We can verify this hypothesis by analysing the details of our mock samples which were built using a HOD model to reproduce the joint distribution of luminosity, colour and clustering amplitude in VIPERS \citep{delatorre13a}.

In the mock catalogues more luminous galaxies tend to be centrals for both blue and red classes.  Consequently the satellite fractions are lower in the volume-limited samples with respect to the flux-limited samples, which is what we were aiming for to minimise the non-linear contribution to the velocity field. This effect is clearly illustrated in Fig. \ref{fig:sat_frac_mocks} which plots the satellite fraction $\mathrm{f_{sat}}$ as a function of redshift in the two cases. A large fraction of the red satellite galaxies are faint and so are effectively removed by the luminosity threshold. However, in absolute terms this fraction is still much higher than for the blue galaxies (40\% vs. 7\%), showing the predominance of red satellites in group environments.  Evidently, to select red galaxies that are mostly centrals, one should select at an even higher luminosity threshold, which would make the sample very sparse and too small for a quantitative analysis. On the other hand, for the blue galaxies, the VIPERS HOD analysis that was used to build the mock samples predicts that a large fraction are central galaxies of their haloes. The difference in $\mathrm{f_{sat}}$ between the flux- and volume-limited samples reduces at high redshifts since the luminosity cut there is closer to the original flux cut of the parent sample.

We then made the assumption that this interpretation applies equally well to the VIPERS data, which is reasonable given that the HOD fit used in the mocks was calibrated using the VIPERS dataset. This choice is further corroborated by more recent HOD analyses that are ongoing with the VIPERS data.

	\begin{figure}
    	\centering
		\includegraphics[scale=0.2]{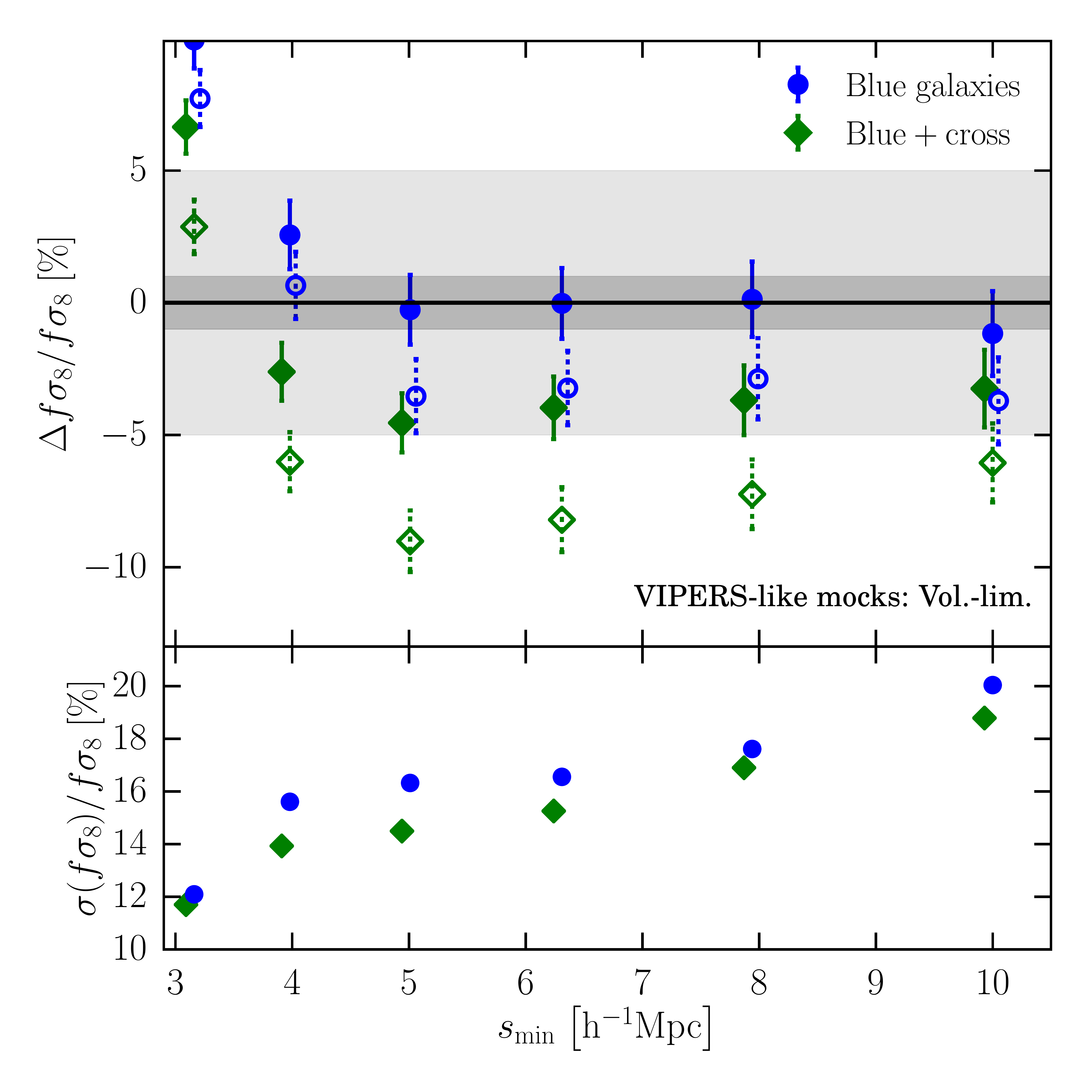}
		\caption{Top panel: same as in Fig. \ref{fig:vol-lim_syst_ideal} but now using the volume-limited samples in fully VIPERS-like mocks. Filled points result from using the Scoccimarro model while the empty markers show the growth rate estimates adopting the dispersion model. Both models adopt a second Gaussian damping factor to mimic the effect of spectroscopic redshift errors in the galaxy redshifts (see Sect. \ref{sec:tests_spec}). The shaded regions are the $5\%$ and $1\%$ intervals around the unbiased case. Bottom panel: relative statistical uncertainties expected for a single VIPERS-like realisation. For clarity, only the Scoccimarro model case is shown.}\label{fig:vol-lim_syst_spec}
	\end{figure}

\subsection{Test on full VIPERS-like mocks}		\label{sec:tests_spec}

Before proceeding to the analysis of the luminous galaxy samples from VIPERS,  we performed a final test of the modelling when all observational effects are included (masks, TSR and redshift measurement errors), analysing the fully realistic VIPERS-like mocks. Following \citet{delatorre16} and \citet{pezzotta16}, redshift errors were accounted for in the RSD modelling through an extra Gaussian damping factor, whose dispersion is fixed to the known rms value of the VIPERS redshift errors. In the mocks, this value is $\sigma_z=4.7\times10^{-4}\(1+z\)$, corresponding to the first estimate from the PDR-1 \citep{Guzzo14}, while when fitting the data we used the most updated estimate from PDR-2 that is $\sigma_z=5.4\times10^{-4}\(1+z\)$ \citep{Scodeggio16}. We also stress that, based on the results of Sect.~\ref{sec:meas}, in all computations (on the mocks, and in the following, on the data), we did not applied the small-scale angular correction using the $\rm{w^A}$ weights.

The results are plotted in Fig. \ref{fig:vol-lim_syst_spec}, where top panel clearly indicates that blue galaxies in the volume-limited sample trace the quasi-linear RSD better than objects in the flux-limited sample, yielding with both models systematic errors within $5\%$ down to very small scales. The Scoccimarro model, in particular, provided virtually unbiased estimates when using scales $s_{\rm{min}}\ge 5\mhmpc$. This further regularisation could be interpreted as due to a further depletion of galaxy pairs with high velocity differences, which cannot be observed after applying SPOC, the VIPERS slit assignment scheme. 

In the same figure we also show the results for the joint fit of the blue auto-correlation and the blue-red cross-correlation, which gave again less satisfactory results. This combination is clearly affected by the poor performances of the red population. The interest in adding this combination was clearly stimulated by the hope to reduce statistical errors. This is in fact the case, as shown the bottom panel of the same figure, where statistical errors expected for a single VIPERS-like realisation using the Scoccimarro model are shown.  However, the gain is marginal when compared to the increase in the systematic bias with respect to the use of blue galaxies alone. 

Based on these results, in the application to the real VIPERS data we will therefore not use red galaxy data (auto- and cross-correlation) as a contribution to our main conclusions regarding $f\sigma_8$. Nevertheless, we will report the measurements for the VIPERS real red galaxy data as it is interesting to see how the consistency of the growth-rate measurements compares with what is seen in the mocks, and discuss this comparison in Appendix~\ref{app:redmodel}.

\section{Results on the VIPERS data}\label{sec:results}
  
Our detailed mock experiments suggest that an analysis of the blue luminous galaxy sample using the Scoccimarro RSD model should allow us to obtain unbiased measurements of $f\sigma_8$ from the data, down to $s=5\mhmpc$ (Fig.~\ref{fig:vol-lim_syst_spec}). But although the mock samples are realistic in many respects, they cannot be expected to match all aspects of the real VIPERS sample. We therefore performed an additional robustness test to check the stability of the values of $f\sigma_8$ measured from the data as a function of $s_{\rm{min}}$, as was done with the mocks.

\begin{figure}
	\centering
		\includegraphics[scale=0.2]{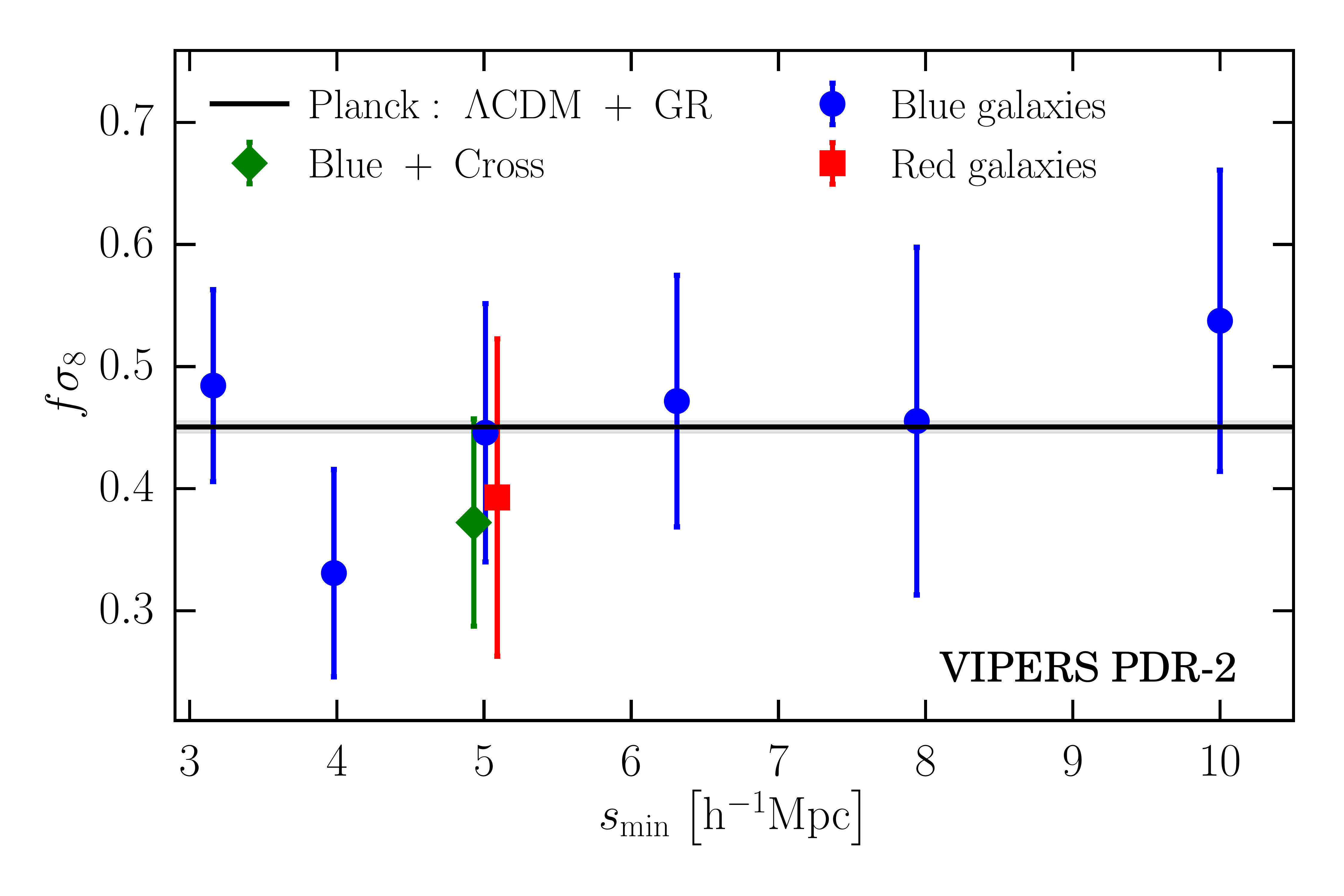}
		\caption{Dependence on the minimum fitted scale, $s_{min}$, of the estimate of $f\sigma_8$ from the VIPERS PDR-2 data, using a volume-limited sample of luminous blue galaxies (blue circles). We also show for comparison, for the reference $s_{min}=5\mhmpc$, estimates based on the auto-correlation of VIPERS luminous red galaxies (red square) and from the blue+cross joint fit (green diamond). All estimates use the Scoccimarro model.} \label{fig:fs8_s-min_data}
	\end{figure}

\begin{figure}
	\centering
		\includegraphics[scale=0.2]{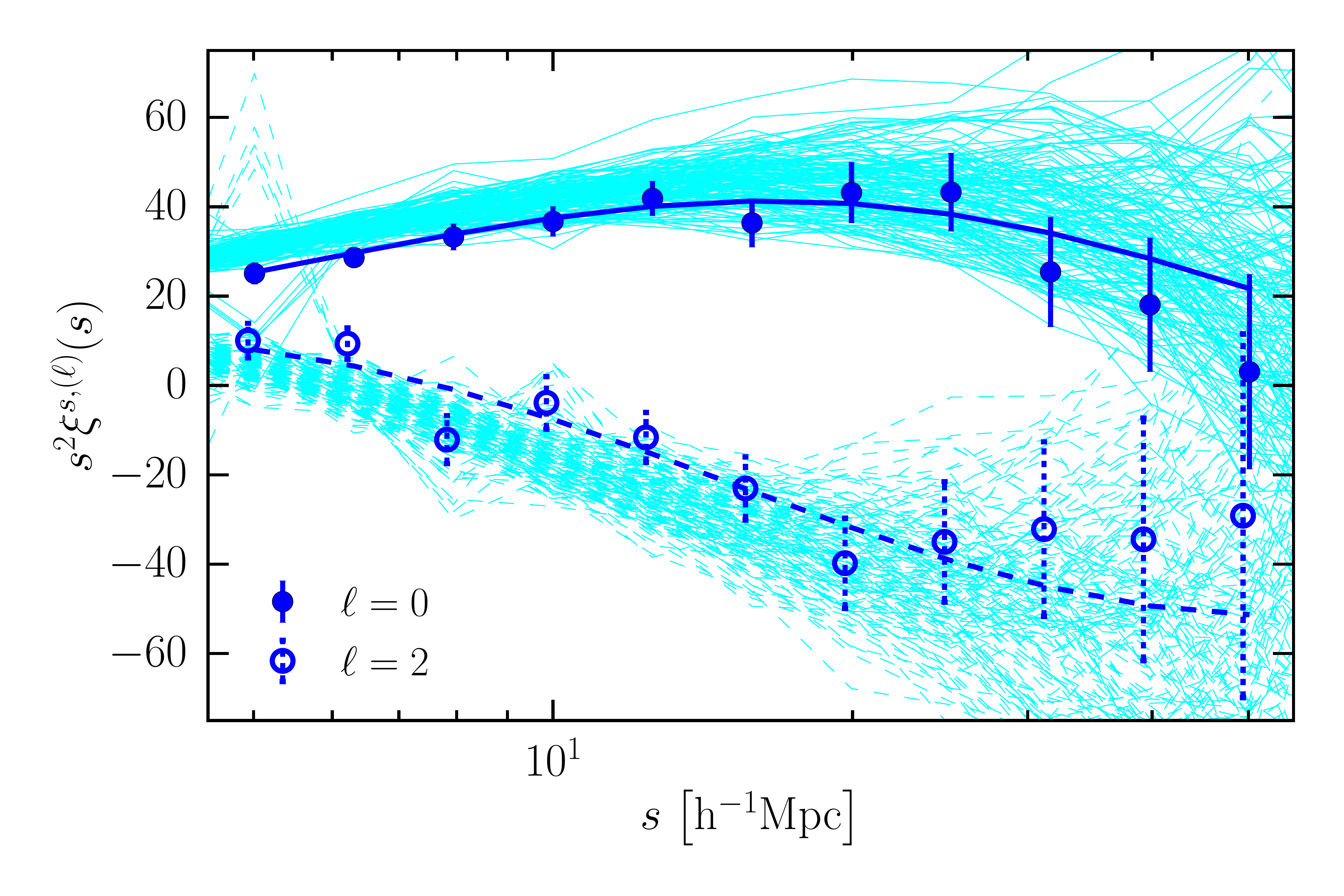}
		\caption{Monopole (filled circles) and quadrupole (empty circles) of the two-point correlation function of VIPERS luminous blue galaxies; the measurements are shown together with the corresponding best-fit model multipoles (thick lines), corresponding to the $s_{min}=5 \mhmpc$ estimate of Fig.~\ref{fig:fs8_s-min_data}. The quadrupole points are slightly shifted along the separation axis for clarity.  The cyan thin continuous and dashed lines show for comparison the corresponding measurements from the 153 mock samples. A version of this plot for the red population is discussed in Appendix~\ref{app:redmodel}.}\label{fig:best-fit_mps}
	\end{figure}

The results are shown in Fig. \ref{fig:fs8_s-min_data}. The recovered values of $f\sigma_8$ are very stable, down to the minimum scales considered, with some oscillation around $s_{\rm{min}}=4\mhmpc$, which we know coincides with the scale where the quadrupole changes sign and is intrinsically difficult to fit. Based on the mock results, however, we preferred to be conservative and excluded from our reference estimate measurements below $s_{\rm{min}}\simeq5\mhmpc$, since this is the range where non-linear effects and non-linear bias might not have been fully captured by our VIPERS mocks. For this same minimum scale $s_{min}$, we have also plotted in the same figure the estimate one obtains by modelling the auto-correlation of the luminous red sample and from a joint fit of the blue auto-correlation and the cross-correlation function. The red galaxy estimate is 13\% lower than the blue galaxy value, which is in agreement with the results of the mock catalogues, but in a less dramatic way (see Fig. \ref{fig:vol-lim_syst_ideal}).
For the chosen reference value of $s_{\rm{min}}$ we plot in Fig.~\ref{fig:best-fit_mps}  the monopole $\(\ell=0\)$ and quadrupole $\(\ell=2\)$ of the VIPERS luminous blue galaxy sample, together with their best fit models. The close agreement between the data and the mocks is impressive -- especially so for the quadrupole, given that the HOD mocks were not built with any requirement to match this moment of the two-point correlation function.  Even though we have used simple probabilistic algorithms to mimic colour sub-classes of galaxies, it seems that the resulting galaxy catalogues are reassuringly realistic in their redshift-space properties.  For comparison, the accuracy of the mock red galaxy sample in reproducing the real data is discussed in Appendix \ref{app:redmodel}.

\begin{figure*}
    	\centering
		\includegraphics[scale=0.35]{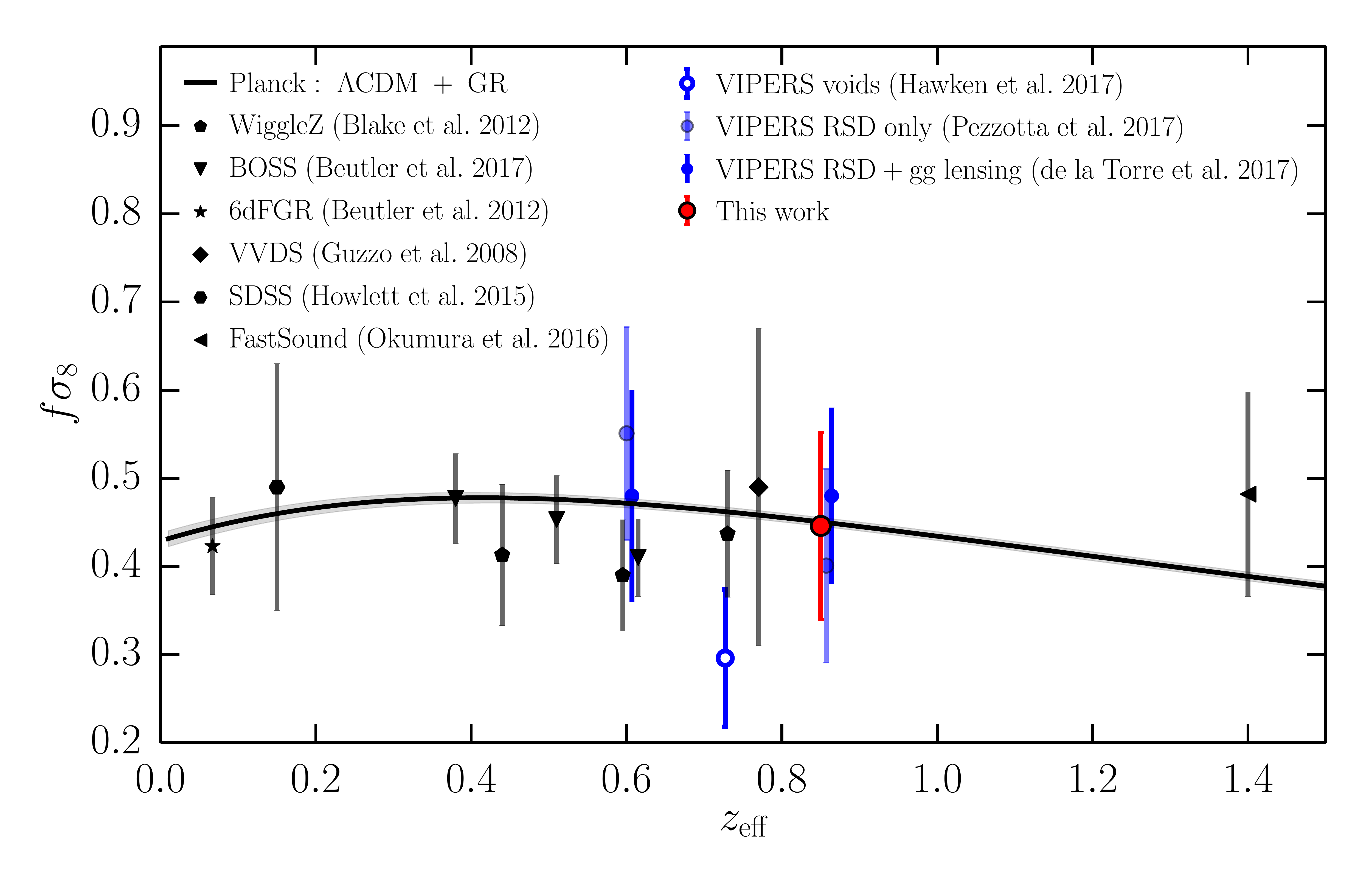}
		\caption{Estimate of the linear growth rate $f\sigma_8$ obtained here from the luminous blue galaxy sample (red circle),  compared to other VIPERS measurements using different techniques, together with those from other surveys. The black solid curve shows the predictions of General Relativity with a $\Lambda$CDM model with parameters set to Planck 2015 \citep{planck15} results and the shaded band corresponds to the related 1-$\sigma$ error.}\label{fig:fs8_surveys}
	\end{figure*}

The results of the joint fit to monopole and quadrupole of the luminous blue galaxies yielded
	\begin{equation}
		f\sigma_8\(z\simeq0.85\) = 0.45\pm0.11 \,\,\,\, ,						\label{eq:fs8-data}
	\end{equation}
which agrees very well with the complementary VIPERS measurements based on the full PDR-2 sample \citep{pezzotta16}, its combination with galaxy-galaxy lensing \citep{delatorre16}, and using galaxy outflows from cosmic voids \citep{hawken16}.  All these measurements are shown in Fig. \ref{fig:fs8_surveys}. Such an agreement provides a posteriori reassurance of the quality of non-linear corrections applied in all these different analyses.  We stress that the sample used here differs significantly from the full-survey samples used in \citet{pezzotta16} and \citet{delatorre16}. In this respect, the measurement based on galaxy outflows from voids \citep{hawken16} is the most novel technique, for which systematic biases are still to be fully explored.  Nevertheless, the value obtained using voids remains within $\sim 2\sigma$ of the other estimates.

In the same figure we also report previous measurements from other large galaxy redshift surveys.  The value obtained here using luminous blue galaxies confirms the agreement of the VIPERS estimates and of virtually all recent surveys with the standard $\Lambda$CDM model with cosmological parameters set to the Planck values (solid line). 
\phantom{\cite{beutler17,blake12,beutler12,okumura16,howlett15}}

\section{Discussion and conclusions}

The relative clustering signal of red and blue galaxies was used in previous works as a way to reduce  statistical errors on the growth rate \citep{blake13,ross14,pearson16} following the idea of \citet{mcdonald09}.  A prerequisite for these multi-tracer analyses is that there is no systematic difference in the inferred growth rate from the individual populations.  This was explicitly verified in the analysis by \citet{ross14} who find that the growth rate measurements from blue and red sub-samples in the CMASS BOSS survey are compatible with each other with a minimum fitting scale of 30 \hmpc.

Pushing the RSD analysis to smaller scales reveals systematic differences in the velocity fields traced by red and blue galaxies \citep{guzzo97}.  In this paper we have taken advantage of these differences in order to minimise the systematic uncertainties in the measurement of the growth rate. We have benefited from the VIPERS broad (essentially flux-limited) selection function, which targeted galaxies uniformly and with high sampling rate, independently of type and colour. We used the rest-frame $UV$ colours to select sub-samples of blue and red galaxies which largely coincide with active and passive galaxy types. We modelled the luminosity evolution of each class to high precision, defining samples that we argue are homogeneously selected in colour and luminosity over the range $0.6 \le z \le 1.0$.   

A clear trend is seen in the data that the red sample has higher velocity dispersion evident by the fingers-of-God feature (See Fig. \ref{fig:xirppi_vol-lim}).  Red galaxies are typically found in high density regions while blue galaxies tend to populate low density environments. This difference is accounted for in the halo occupation distribution (HOD) framework, in which the  red sample has a higher satellite fraction characteristic of massive dark matter haloes, while the blue sample is predominantly made up of central galaxies in lower mass haloes.  Guided by this idea, we have explored the possibility of selecting blue objects as preferential RSD tracers in the linear regime. We tested this hypothesis using a set of VIPERS mock catalogues that were built within the same HOD theoretical framework.

The main findings of the analysis on mock catalogues is that the accuracy of the growth-rate $f\sigma_8$ measurement can be improved by using sub-samples of galaxies.  Blue galaxies, which are more likely to be central galaxies in low mass haloes, give less biased measurements of $f\sigma_8$.  Furthermore, by constructing volume limited samples, and in general by applying a luminosity cut, we can reduce the satellite fraction and further reduce the non-linear motions.

The mock experiments additionally show that the auto-correlation of the volume-limited blue galaxy sample provides the best estimate of the growth rate while maintaining a small statistical error.  In fact, the statistical error we find matches that for the full galaxy sample \citep{pezzotta16} despite using a sample that is half the size.  This is consistent with the expectations from \citet{bianchi12}, who show that the error is driven by volume and scales weakly with number density.  An additional consideration that can affect the error is the number of free parameters in the model. In the full analysis of \citet{pezzotta16} the parameters $\sigma_8$ and $f$ were allowed to vary independently.  This freedom is needed in the \citet{taruya10} parametrization but not for the \citet{scoccimarro04} model adopted here.  As long as a good fit can be found, reducing the number of model degrees of freedom should lead to stronger constraints. In making this argument, we do have to be aware of potential biases from effects that are omitted from both mocks and theory; in that case, recovery of the true growth rate from simulated data need not imply that unbiased results will also result from the same method applied to actual observations. Indeed, \citet{reid14} pointed out that different models of aligned peculiar velocities of central galaxies in massive structures may have significant impact on the RSD signal at separations of a few Mpc. The corrections seen by Reid et al. are not significant in comparison with the random errors in the current VIPERS work, which is why we have been content to construct mocks using the simple assumption that central galaxies sit  at rest in their haloes. But the challenge of constructing full realistic mocks will only increase as RSD studies move into regimes of higher precision.

Applying our current methodology to the VIPERS data, we produced a new measurement of the growth rate at the effective redshift $z=0.85$ using the auto-correlation of the volume-limited blue galaxy sample alone: $f\sigma_8(z=0.85)=0.45\pm0.11$.  The estimate is consistent at the $1\text{-}\sigma$ level with those obtained from the full galaxy sample from the VIPERS dataset, and is fully consistent with the $\Lambda$CDM model constrained by the Planck data.

This work demonstrates that improved accuracy in RSD measurements can be achieved by identifying appropriate galaxy tracers. By selecting particular galaxy samples by colour and luminosity we are able to use relatively straightforward theoretical models without compromising the systematic and statistical precision on the estimates of the cosmological parameters. Furthermore, the use of this technique helps us in pushing the statistical analysis of RSD to relatively lower scales where the clustering signal is measured with higher statistical precision. We look forward to seeing this robust technique applied to forthcoming large surveys where the statistical precision on the growth rate is capable of reaching the few percent level.

\begin{acknowledgements}

We acknowledge the crucial contribution of the ESO staff for the
management of service observations. In particular, we are deeply
grateful to M. Hilker for his constant help and support of this
programme. Italian participation to VIPERS has been funded by INAF
through PRIN 2008, 2010, 2014 and 2015 programs. LG, FGM,BRG and JB
acknowledge support from the European Research Council through grant
n.~291521. OLF acknowledges support from the European Research Council
through grant n.~268107. JAP acknowledges support of the European
Research Council through the COSFORM ERC Advanced Research Grant (\#
670193). GDL acknowledges financial support from the European Research
Council through grant n.~202781. RT acknowledges financial support
from the European Research Council through grant n.~202686. AP, KM,
and JK have been supported by the National Science Centre (grants
UMO-2012/07/B/ST9/04425 and UMO-2013/09/D/ST9/04030). EB, FM and LM
acknowledge the support from grants ASI-INAF I/023/12/0 and PRIN MIUR
2010-2011. LM also acknowledges financial support from PRIN INAF
2012. SDLT and MP acknowledge the support of the OCEVU Labex
(ANR-11-LABX-0060) and the A*MIDEX project (ANR-11-IDEX-0001-02)
funded by the "Investissements d'Avenir" French government programme
managed by the ANR. TM and SA acknowledge financial
support from the ANR Spin(e) through the French grant
ANR-13-BS05-0005.

The Big MultiDark Database used in this paper and the web application providing online access to it were constructed as part of the activities of the German Astrophysical Virtual Observatory as result of a collaboration between the Leibniz-Institute for Astrophysics Potsdam (AIP) and the Spanish MultiDark Consolider Project CSD2009-00064. The Bolshoi and MultiDark simulations were run on the NASA's Pleiades supercomputer at the NASA Ames Research Center.

We thank the reviewer for the insightful comments that led us to improve the overall message of this work.

\end{acknowledgements}

\bibliographystyle{aa}
\bibliography{biblio_gg,biblio_VIPERS_v4}

\appendix

\section{Angular weights}\label{ap:weights}

The instrumental constraints of placing slits on the galaxies imposes a systematic exclusion of pairs with small angular separations.  The effect is important on scales $<1\mhmpc$ in VIPERS \citep{delatorre13a}. To account for the missing power, we weighted galaxy pairs according to the angular separation with the function $\rm{w^{A}}\(\theta\)$ \citep{hawkins03} defined as
    \begin{equation}
        \frac{1}{\rm{w^{A}}\(\theta\)}=\frac{1+w_{\rm{s}}\(\theta\)}{1+w_{\rm{p}}\(\theta\)}\; ,       \label{eq:ang-weights}
    \end{equation}
where $w_{\rm{s}}\(\theta\)$ and $w_{\rm{p}}\(\theta\)$ are the angular two-point correlation function of the spectroscopically observed and underlying parent samples of galaxies and $\theta$ is the pair separation angle.

\begin{figure}
        \centering
        \includegraphics[scale=0.2]{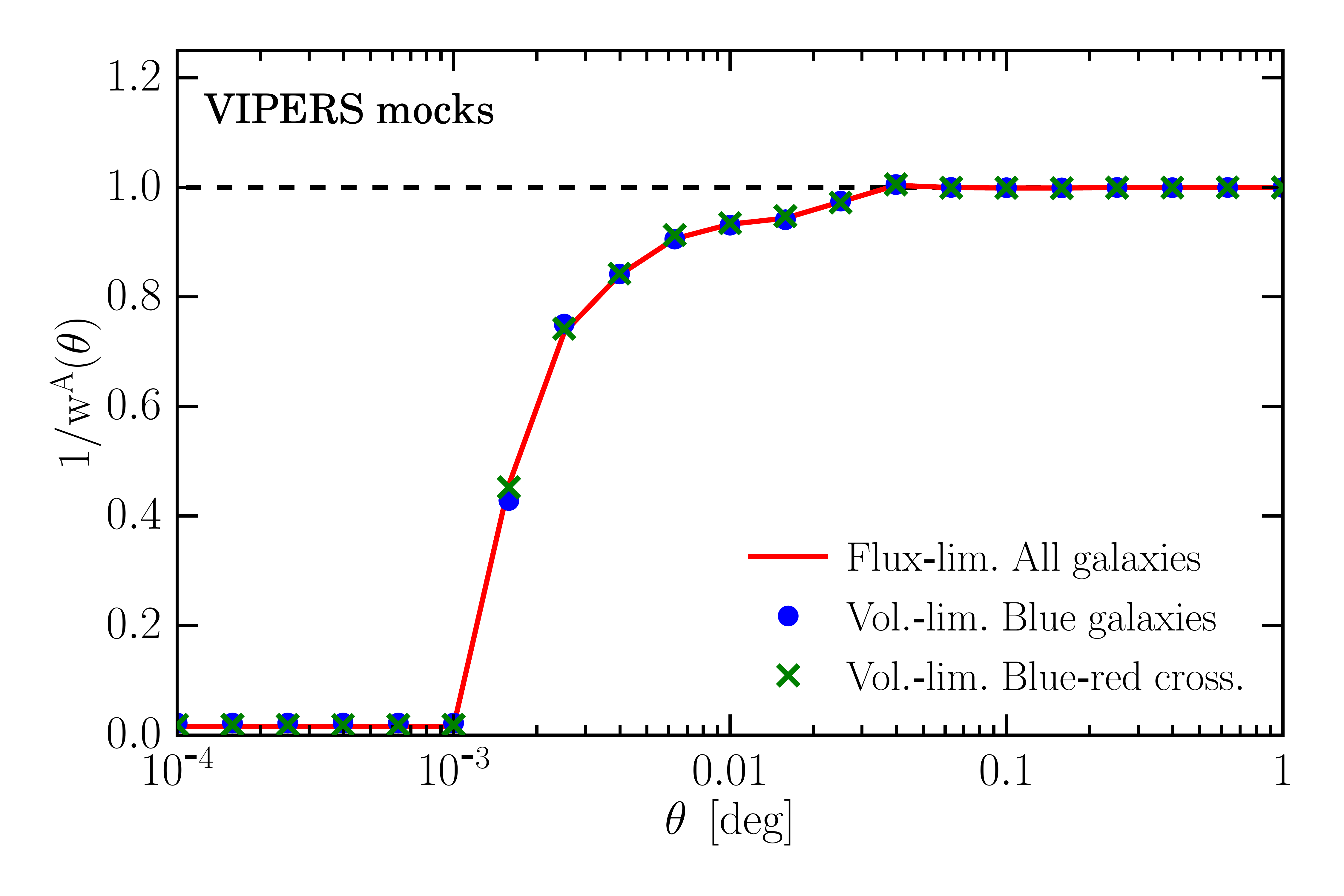}
        \caption{Angular completeness $1/\rm{w^A}\(\theta\)$ as a function of the pair separation angle $\theta$ estimated using the VIPERS mocks. All measurements are performed over the  redshift range $0.6\le z\le1.0$. The auto-correlation statistic of the full magnitude-limited sample (red continuous line) provides estimates of $\rm{w^A}$ almost identical to those obtained using auto-correlation of volume-limited samples of blue galaxies (blue dots) and their cross correlation (green crosses) with the red ones. 
        }\label{fig:ang_corr}
    \end{figure}

As with the target sampling rate (Sect. \ref{incompleteness}), the application of the angular weight requires having the photometric parent sample in hand.  However, using mock catalogues we demonstrated that the weight does not depend on the sample (see Fig. \ref{fig:ang_corr}).  Thus we applied the angular weights computed for the flux-limited sample.

The effect of the angular weights together with the sampling weights is shown in Fig. \ref{fig:xi_cc_corr}.  We see that the target sampling rate produces the largest systematic effect.  The angular correction does not further improve the measurement. For this reason we decided to not to use this correction when analysing the VIPERS data.

\section{Impact of smoothing the data covariance matrix on the estimates of $f\sigma_8$}\label{app:smoothing}

\begin{figure}
	\centering
		\includegraphics[scale=0.2]{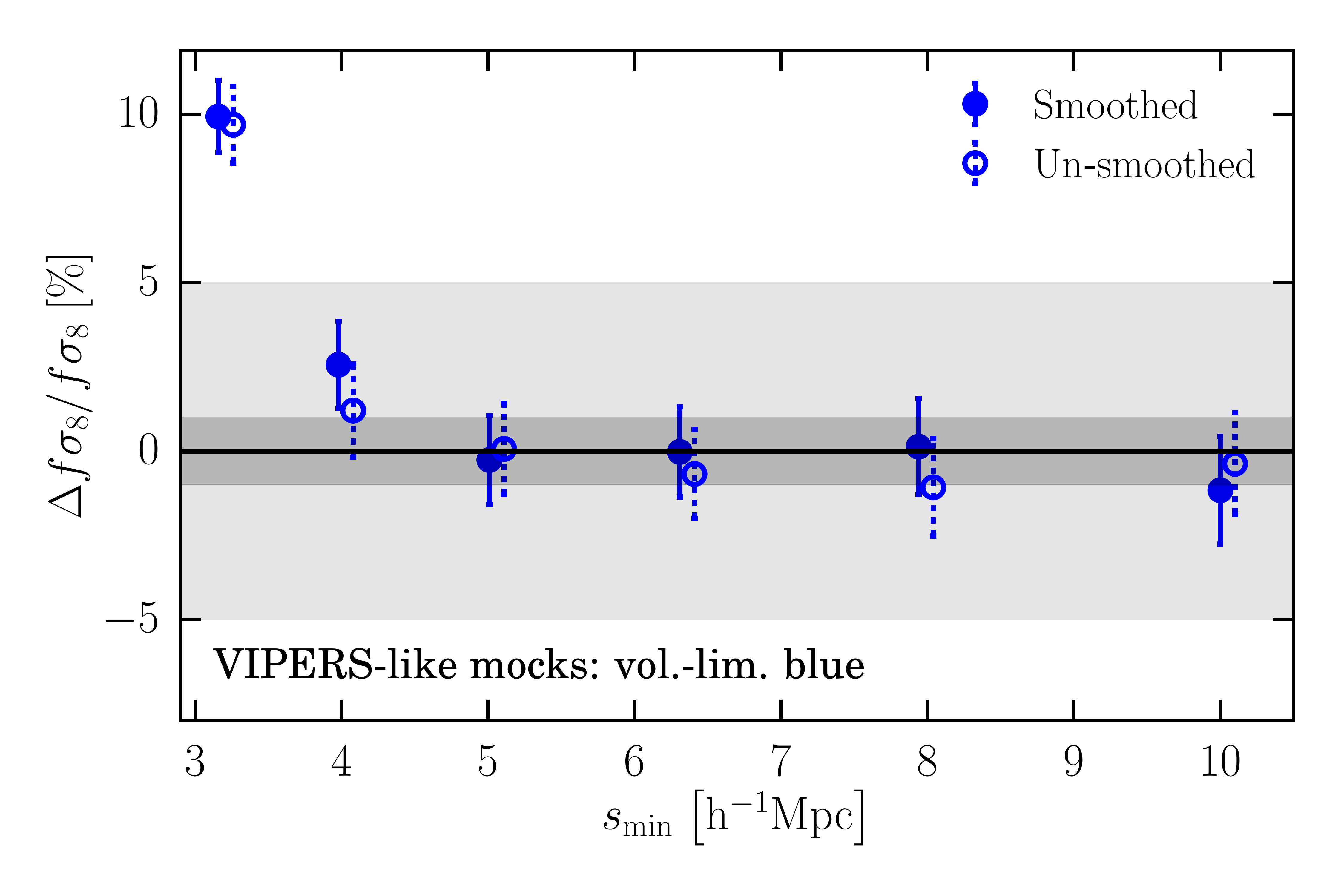}
		\caption{Impact of the smoothing of the data covariance matrix on the systematic errors on $f\sigma_8$ (filled circles, same as the filled blue circles in Fig. \ref{fig:vol-lim_syst_spec}) when fitting the auto-correlation function of luminous blue galaxies in VIPERS-like mock samples. Measurements obtained using the raw estimates of the data covariance matrix are also shown as empty circles. In both cases error bars give the errors on the mean of the 153 mock samples. Fits are performed using the Scoccimarro model.}\label{fig:smooth-raw_fs8}
	\end{figure}

In Sect. \ref{sec:cov} we used the boxcar algorithm \citep{mandelbaum13} to smooth the off-diagonal elements of the data covariance matrix. The visual analysis of Fig. \ref{fig:corr_ac+cc_data} shows that the smoothing does not alter the global structure of the correlation (or equivalently covariance) matrix. When smoothing the covariance matrix, it is crucial that such a scheme does not alter the estimated values of the fitting parameters. In Fig. \ref{fig:smooth-raw_fs8} we show the impact of the smoothing of data covariance matrix on the measurements of the linear growth rate $f\sigma_8$ with respect to the use of the raw estimates of the covariance matrix. It is clear that the smoothing scheme does not have a statistically significant effect on the best estimates of $f\sigma_8$ while it only makes the results more stable in the range of minimum fitting scales of interest, which is $5\mhmpc\lesssim s_{\rm{min}}\lesssim8\mhmpc$.

\section{Dependence of results on the fitting method}\label{app:fit}

\begin{figure}
\centering
	\includegraphics[scale=0.6]{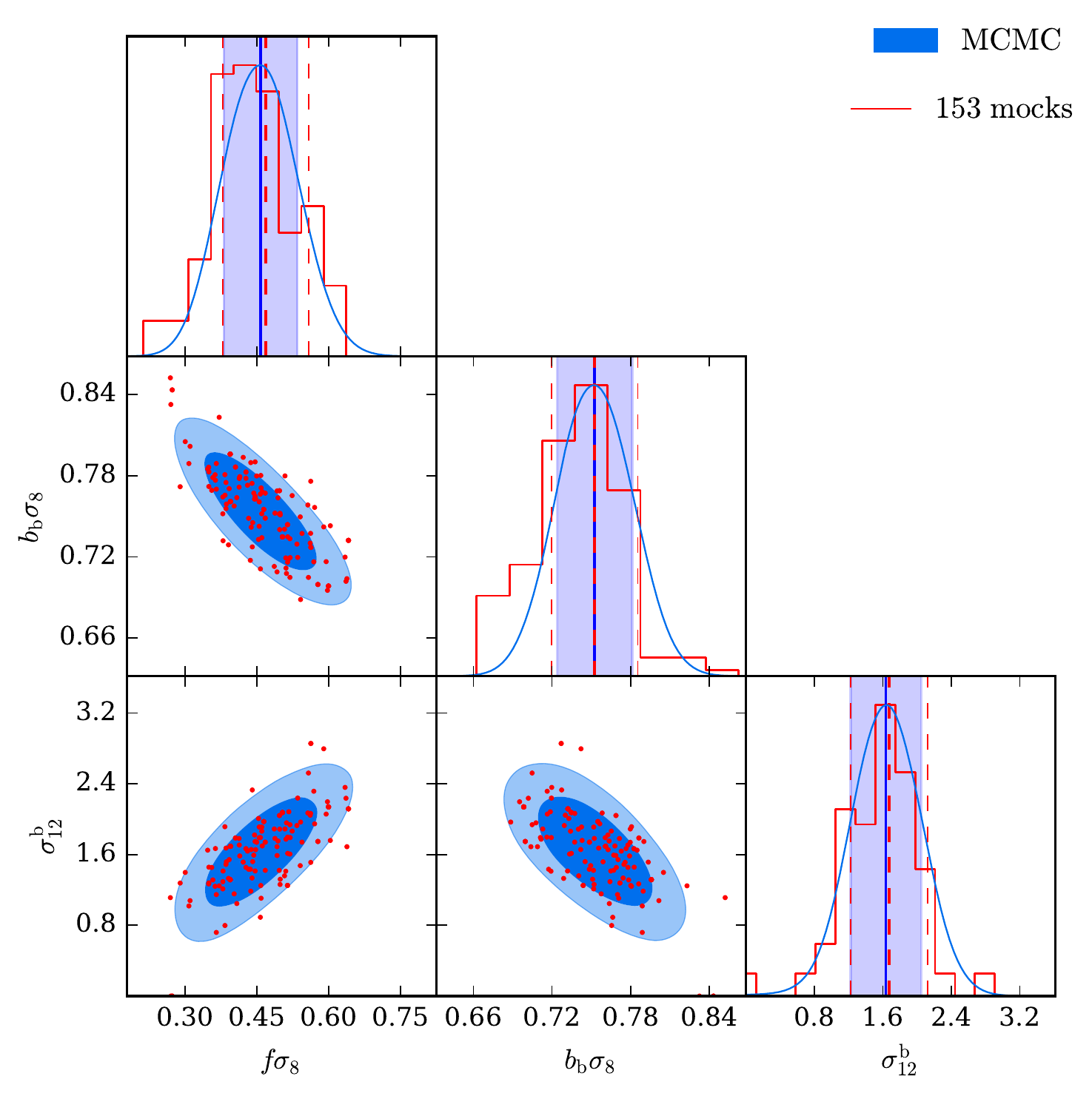}
		\caption{Results of fitting the auto-correlation of blue galaxies in the volume-limited VIPERS-like mocks between $5\mhmpc<s<50\mhmpc$ with the Scoccimarro model. In all panels, blue colour shows results using the MCMC technique while red colour represents results obtained fitting single mocks. The blue smooth curves and coloured contours show respectively the one- and two-dimensional marginalised posterior likelihood distributions obtained running an MCMC algorithm. The red points and the histograms correspond to the best-fit values obtained from each of the 153 VIPERS mocks. For each single-parameter distribution, the vertical thick solid blue and red dashed lines show the best-fit values, while the vertical shaded blue stripes and thin dashed red lines give the corresponding 1-$\sigma$ statistical errors, respectively.}\label{fig:mcmc-vs-mocks}
	\end{figure}

In Sect. \ref{sec:tests} we explored the parameter space through a Monte Carlo Markov Chain (MCMC) technique where the measured multipole moments $y^{s,(0)}(s)$ and $y^{s,(2)}(s)$ averaged over 153 mock samples were fitted with the theoretical models using the related covariance matrix. The best-fit estimate for a given fitting parameter was taken as its value corresponding to the maximum of the one-dimensional marginalised posterior likelihood with the statistical error given by the corresponding $68\%$ confidence level. But ideally one would like to apply the same procedure, adopted in the case of real data, to the measurements from each mock and take the mean of the best-fit estimates as the best estimate of the fitting parameter with its dispersion among 153 mocks as the statistical 1-$\sigma$ error. However, the latter method is computationally time consuming given the large set of cases we explored in this work, in terms of different theoretical models, fitting scales and tracers. Here we show that these methods both yield to the same results in terms of the best estimate and statistical errors of the fitting parameters. In order to fairly compare the two fitting methods, when using the MCMC technique, here we fit the mean estimate of the multipole moments among 153 mocks with the data covariance matrix related to a single realization. The results are shown in Fig. \ref{fig:mcmc-vs-mocks}. In particular, the distribution of the best fit values among 153 mock realisations qualitatively agrees with the 68$\%$ and 95$\%$ confidence levels of the two-dimensional marginalised posterior likelihoods obtained using the MCMC technique in the 2D plots. Furthermore, the marginalised one-dimensional posterior provides a good match to the frequency histograms. Also the best-fit estimates and the related 1-$\sigma$ statistical errors obtained from the two methods are in a very good agreement. Overall we conclude that the results obtained in this work do not depend on the specific technique used here. In Sect. \ref{sec:tests} we therefore used the Monte Carlo approach to fit the mean estimates of the 2PCF multipoles among our 153 mocks with the related covariance matrix.

\section{Errors on the number counts} \label{app:err_pred}

The covariance matrix of the number counts is computed assuming a linear bias relation between galaxies and the overall matter distribution \citep[see e.g.][]{crocce10, hoffmann15}. The number count of galaxies $N_i$ in bin $i$ (e.g. a bin of $UV$ colour) measured in a volume $V$ centred on position $\mathbf{r}$ can be written as
	\begin{equation}
		N_i\(\mathbf{r}\)=\bar{N}_i\[1+\delta_i\(\mathbf{r}\)\]+\delta N_i^{\rm{sn}}\; .							\label{eq:ep0}
	\end{equation}
The mean number count in the bin is $\bar{N}_i$, $\delta_i$ is the mean density contrast of the galaxy field in the volume and $\delta N_i^{\rm{sn}}$ is the noise which we assume to be Poissonian.

In the approximation of the linear, local and scale-independent galaxy bias $b_i$, we can rewrite Eq. \eqref{eq:ep0} as
	\begin{equation}
		N_i\(\mathbf{r}\)=\bar{N}_i\[1+b_i\delta_m\(\mathbf{r}\)\]+\delta N_i^{\rm{sn}}\; ,						\label{eq:ep1}
	\end{equation}
where $\delta_m$ is the density contrast related to the overall matter in the volume $V$ around $\mathbf{r}$. The covariance matrix $C_{ij}$ of the number counts in the measurement bins $i$ and $j$ is defined as
	\begin{equation}
		C_{ij} = \langle \Delta_i\Delta_j\rangle=\frac{1}{N_s}\sum_{k=1}^{N_s}\Delta_i^k\Delta_j^k\; ,																\label{eq:ep3}
	\end{equation}
where $N_s$ is the number of independent realisations and $\smash{\Delta_i^k=N_i^k-\bar{N}_i}$ is the difference between the measurement in bin $i$ drawn from the $k^{\rm th}$ realisation and the corresponding mean value. The covariance matrix in Eq. \eqref{eq:ep3} results from a combination of two terms
	\begin{equation}
		C_{ij}=C_{ij}^{\rm{SV}}+C_{ij}^{\rm{SN}}\; .																	\label{eq:ep4}
	\end{equation}
In Eq. \eqref{eq:ep4} $C_{ij}^{\rm{SV}}$ is the contribution from the sample variance, 
	\begin{equation}
		C_{ij}^{\rm{SV}} = \bar{N}_i\bar{N}_jb_ib_j\sigma_m^2\(V\)\; ,								\label{eq:ep5}
	\end{equation}
where $\sigma_m^2$ is the variance of the matter density contrast in the volume $V$. Having at disposal only one realisation of our dataset, we replace $\bar{N}_i$ with the value directly measured from data.

The second term $\smash{C_{ij}^{\rm{SN}}}$ in Eq. \eqref{eq:ep4} is related to the noise on the measurements. This is assumed to be Poissonian and thus independent of the position $\mathbf{r}$. The Poissonian shot-noise contribution to $C_{ij}$ is diagonal and given by
	\begin{equation}
		C_{ii}^{\rm{SN}}=\bar{N}_i\; . \label{eq:ep6}
	\end{equation}
    
The estimation of $\sigma_m^2\(V\)$ requires the knowledge of the window function which describes the peculiar geometry of the survey. Alternatively, rather than computing directly the window function, one can use a catalogue of randomly distributed points within the survey, with the same radial and angular selection function of the galaxy sample under consideration, to probe the related volume. Another ingredient to compute the variance $\sigma_m^2\(V\)$ is the matter two-point correlation function $\xi_m$ which we compute using the publicly available CAMB code \citep{lewis00} for the reference cosmological model. Once the volume $V$ is populated with a sufficiently dense catalogue of $N_{\rm R}$ random points, the variance $\sigma_m^2\(V\)$ can be computed as
	\begin{equation}
		\sigma_m^2\(V\)=\frac{1}{V^2}\sum_{i'}^{N_{\rm R}}\sum_{j'>i'}^{N_{\rm R}}\xi_m\(|\mathbf{r}_{j'}-\mathbf{r}_{i'}|\)\Delta V_{i'}\Delta V_{j'}\; ,			\label{eq:ep7}
	\end{equation}
with $\Delta V_{i'}$ being the volume probed by the $i'$-th random particle. We approximate $\Delta V_{i'}$ with
	\begin{equation}
		\Delta V = \Delta V_{i'} = \frac{V}{N_{\rm R}}\; .							\label{eq:eq8}
	\end{equation}
The error covariance matrix in Eq. \eqref{eq:ep4} then can be written as
	\begin{equation}
		C_{ij} = \bar{N}_i\bar{N}_jb_ib_j\frac{1}{N_{\rm R}^2}\sum_i^{N_{\rm R}}\sum_{j>i}^{N_{\rm R}}\xi_m\(|\mathbf{r}_j-\mathbf{r}_i|\)+\bar{N}_i\delta^K_{ij}\; ,																	\label{eq:ep9}
	\end{equation}
where the shot noise term is written with the Kronecker delta $\smash{\delta^K_{ij}}$. 
The biases $b_i$ and $b_j$ of galaxies in bins $i$ and $j$ are, in practice difficult to infer directly from data. We thus approximate them with a constant bias value.

\section{Performances of the Scoccimarro model for simulated and real red galaxies}\label{app:redmodel}
Our measurements from the mock samples show that red galaxies alone gave rather biased estimates of the growth rate when using the Scoccimarro model (see Sect. \ref{sec:fit_ideal}). However, the application of the same model to the real red galaxy data yielded a value that, still underestimated, is much closer to the one obtained from the blue galaxies. In this Appendix we investigate the issue in greater detail and discuss the extent to which the mocks are representative of the VIPERS sample.
\begin{figure}
	\centering
	\includegraphics[scale=0.2]{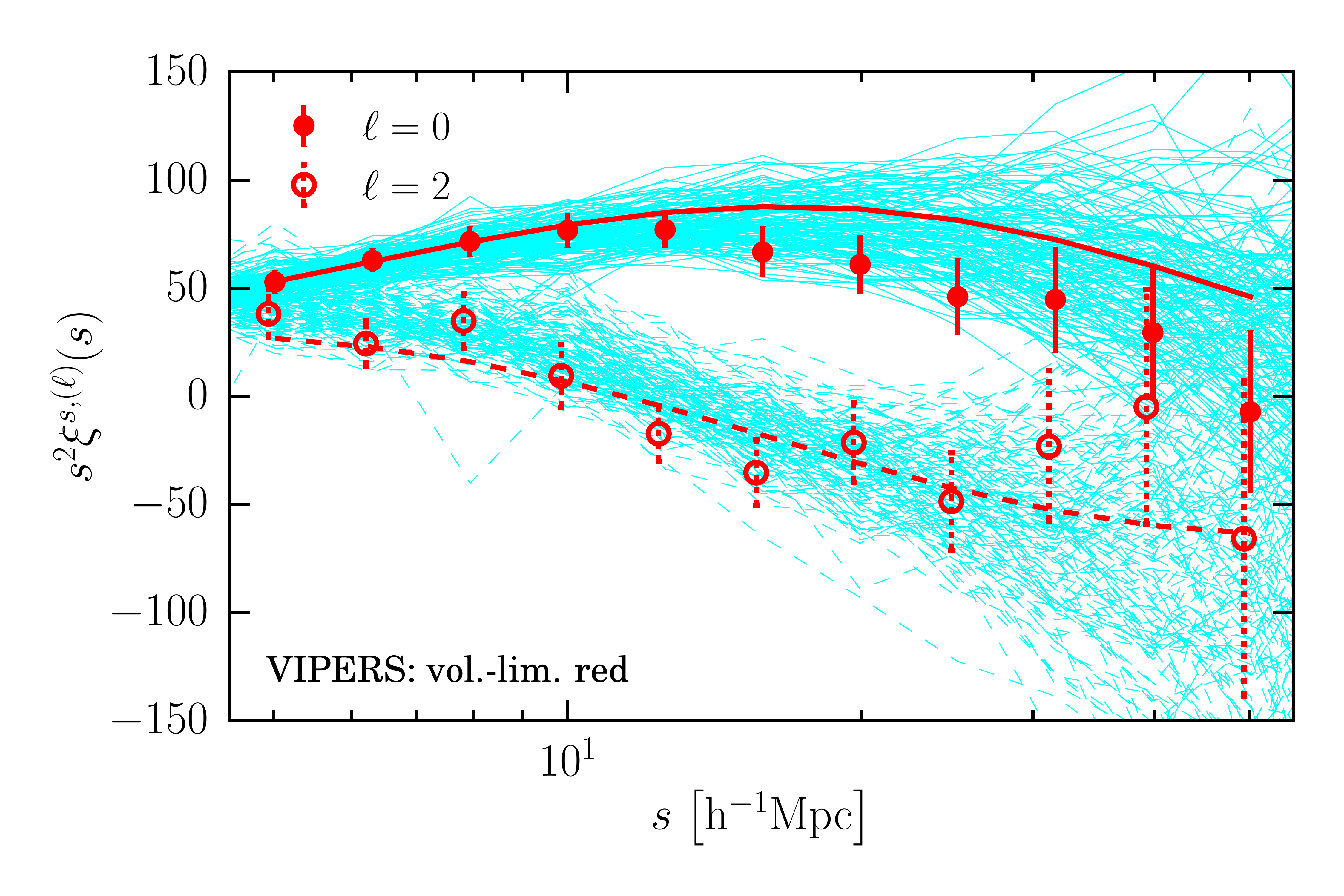}
    \caption{Same as in Fig. \ref{fig:best-fit_mps} but here for the volume-limited samples of red galaxies in VIPERS.} \label{fig:best-fit_mps_red}
\end{figure}
\begin{figure}
	\centering
	\includegraphics[scale=0.2]{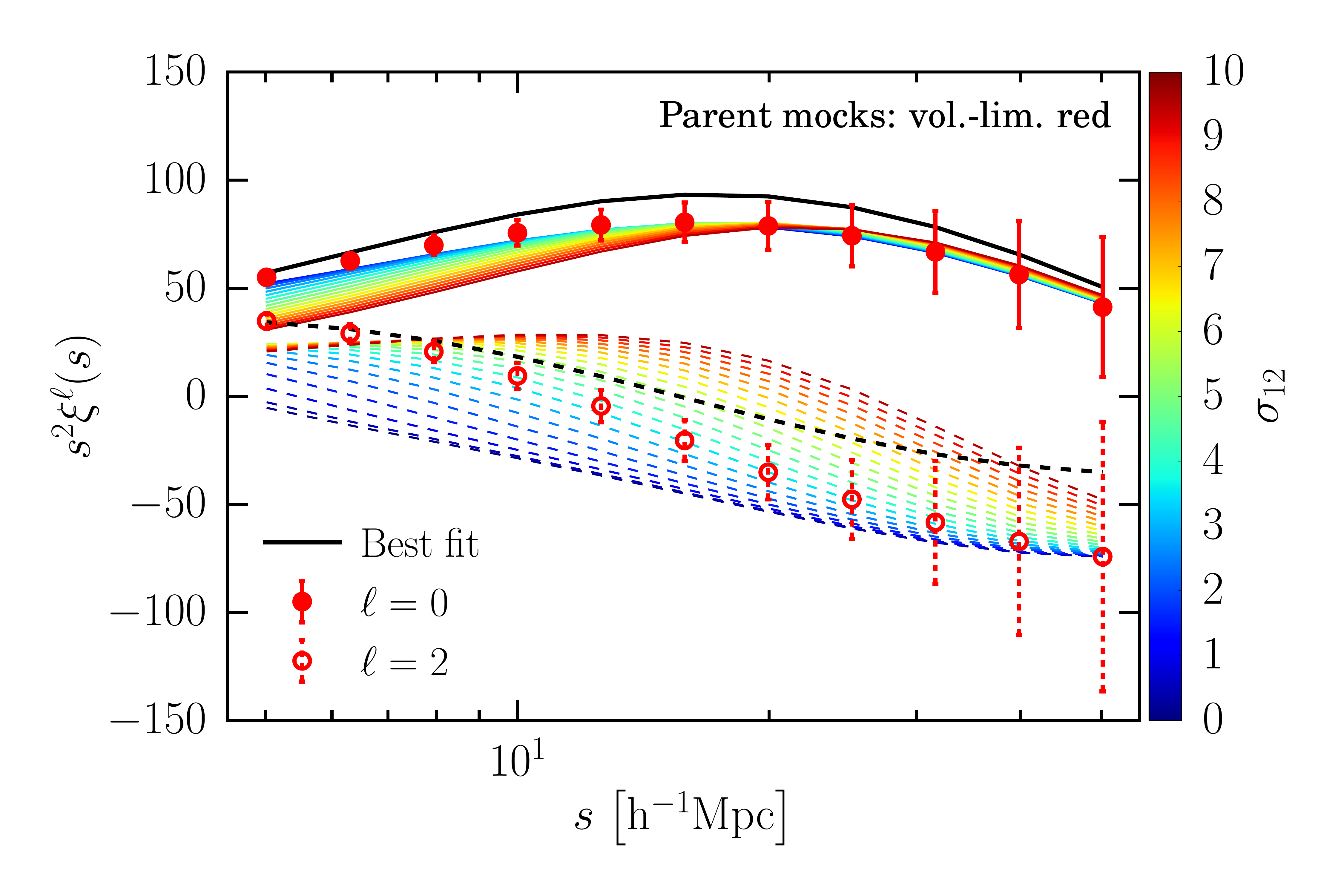}
    \caption{Behaviour of the Scoccimarro model when varying the dispersion parameter $\sigma_{12}$. Points with error bars show the monopole ($\ell=0$) and quadrupole ($\ell=2$) measured using the volume-limited ideal parent mock samples of red galaxies. We show the errors corresponding to a single realisation to highlight the relative difference of precision at different scales as the errors on the mean are too small to appreciate their scale dependence. Black continuous and dashed thick lines show the best-fit  Scoccimarro model with $s_{\rm{min}}=5\mhmpc$. Thin continuous and dashed lines with colour spanning from blue to red are obtained fixing the growth rate to its fiducial value $f\sigma_8=0.46$ and the dispersion parameter varying from $\sigma_{12}=0\mhmpc$ to $\sigma_{12}=10\mhmpc$. The linear bias parameter is set to match the amplitude of the monopole at large scales.}\label{fig:scocc_test}
\end{figure}
Figure~\ref{fig:best-fit_mps_red} is the analogue of Fig.~\ref{fig:best-fit_mps}, showing the monopole and quadrupole of the correlation function of the VIPERS red galaxy sample, compared to those from the ensemble of red galaxy mocks; both refer to volume-limited samples as defined in the main text.  The Scoccimarro model best fit to the data points is overplotted, using $s_{\rm{min}}=5\mhmpc$ as done for the blue galaxy data.  This corresponds to $f\sigma_8\(z=0.84\) = 0.39\pm0.13$, which is 13\% lower than the corresponding value obtained fitting the blue population. This difference, while still indicating the tendency of the red population to deliver lower values of the growth rate, is significantly smaller than what the mocks experiments indicated. It is interesting to understand why, given that the overall agreement of the mocks with the data shown in the figure would seem qualitatively good (in particular once a small difference in the linear bias between the red data and mocks is taken into account). 

Let us then look in more detail into the measurements on the volume-limited samples obtained from the ideal parent mocks, which were used to produce Fig.~\ref{fig:vol-lim_syst_ideal}. Figure \ref{fig:scocc_test} shows the mean monopole and quadrupole of the correlation function. We stress here that these are precise values obtained by averaging the 153 mock realisations, since the aim of our mock experiments is to evidence systematic limitations in the accuracy of the applied model. The best-fit model to these mean measurements for $s_{\rm{min}}=5\mhmpc$ is also shown in Fig.~\ref{fig:scocc_test} that corresponds to the following set of parameters: $f\sigma_8=0.22\left(\pm0.01\right)$, $b_{\rm{r}}\sigma_8=1.208\left(\pm0.002\right)$ and $\sigma_{12}=3.29\left(\pm0.02\right)$.  As extensively discussed in the main text, the recovered $f\sigma_8$ is significantly biased low, compared to the fiducial value of the mocks, $f\sigma_8=0.46$, corresponding to a $-52\%$ systematic error. The figure shows  more clearly what happens: the fit is substantially driven by the points at small separations ($s<8\mhmpc$), due to their small error bars, and the same model cannot fit both small and large scales adequately.  We explore the sensitivity to the different parameters by overplotting a range of models when we fixed the growth rate to the fiducial value, $f\sigma_8=0.46$, and varied the pairwise dispersion $\sigma_{12}$ over the range $[0,10] \mhmpc$; we also set the bias to $b_{\rm{r}}\sigma_8=1.0$, to match the monopole at large scales.  What the curves show is that when the fiducial growth rate is fixed, the model can reproduce the measured monopole with a virtually null pairwise dispersion, but it is not capable of fitting the quadrupole particularly at small scales.  We also note from Fig.~\ref{fig:vol-lim_syst_ideal} that if separations smaller than $s_{\rm{min}}=10\mhmpc$ are excluded, then the systematic error estimated from the red mocks volume-limited sample becomes negligible (the full magnitude-limited sample still gives a large systematic even for such large values of $s_{\rm{min}}$ as shown in Fig.~\ref{fig:flux-lim_syst_ideal_z_0.6-1.0}).

The conclusion is that the higher non-linear velocity component in the mock samples, which is stronger than in the real data (see Fig.~\ref{fig:xirppi_vol-lim}) exacerbates the limitations of the model, evidencing where its weaknesses lie.  The increasingly better performances obtained with the volume-limited sample of real red galaxy data (which clearly show a smaller contribution from nonlinear motions than the mocks) and the blue galaxy experiments are all consistent with the main result of this paper:  that relatively simple RSD models can be applied including measurements down to scales as small as $5\mhmpc$ to obtain virtually unbiased results, provided that the non-linear small-scale component of the velocity field is minimised.  We have shown that an effective way to obtain this is to work with volume-limited samples of blue luminous galaxies.

\end{document}